\pdfoutput=1
\documentclass[sigconf]{acmart}
\AtBeginDocument{%
  \providecommand\BibTeX{{%
    \normalfont B\kern-0.5em{\scshape i\kern-0.25em b}\kern-0.8em\TeX}}}

\acmConference[EDBT/ICDT 2021]{
EDBT 2021: 24th International Conference on Extending Database Technology 
}{March 23-26, 2022}{Nicosia, Cyprus}
\acmBooktitle{EDBT 2021: 24th International Conference on Extending Database Technology 
March 23-26, 2022 Nicosia, Cyprus}
\acmPrice{15.00}
\acmISBN{978-1-4503-XXXX-X/18/06}

\usepackage{amsmath,amsfonts}
\usepackage{graphicx}
\usepackage{textcomp}
\usepackage{xcolor}

\newcommand{\squishlist}{
 \begin{list}{$\bullet$}
  { \setlength{\itemsep}{0pt}
     \setlength{\parsep}{3pt}
     \setlength{\topsep}{3pt}
     \setlength{\partopsep}{0pt}
     \setlength{\leftmargin}{1.5em}
     \setlength{\labelwidth}{1em}
     \setlength{\labelsep}{0.5em} } }

\newcommand{\squishlisttwo}{
 \begin{list}{$\bullet$}
  { \setlength{\itemsep}{0pt}
    \setlength{\parsep}{0pt}
    \setlength{	opsep}{0pt}
    \setlength{\partopsep}{0pt}
    \setlength{\leftmargin}{2em}
    \setlength{\labelwidth}{1.5em}
    \setlength{\labelsep}{0.5em} } }

\newcommand{\squishend}{
  \end{list}  }

\usepackage{graphicx}

\usepackage[font=footnotesize,labelfont=bf]{caption}
\usepackage[ruled,linesnumbered]{algorithm2e}
\usepackage[noend]{algpseudocode}
\usepackage[linguistics]{forest}
 \usetikzlibrary{patterns}
\usepackage{pgfplots}
\usepackage{array}
\usepackage{subcaption}
\usepackage{qtree}
\usepackage{newfloat}
\usepackage{siunitx}
\usepackage{booktabs}
\usepackage{tabularx}

\newcommand{\RED}[1]{  {\color{red}{#1}}}

\newcommand{\silence}[1]{}

\newcommand{\galexiou}[1]{\textcolor{yellow}{#1}}

\newcommand{\framework}{\mbox{QueryER}\xspace} %new index name
\renewcommand\footnotetextcopyrightpermission[1]{}

\usepackage{titlesec}
\titlespacing*{\section}
{.5pt}{2ex }{0ex }
\titlespacing*{\subsection}
{.5pt}{2ex}{0ex}
\titlespacing*{\subsubsection}
{.5pt}{0.4ex}{0ex}
\titleformat{\section}
  {\normalfont\Large\bfseries}
  {\thesection}
  {.5em}
  {\MakeUppercase}
\usepackage{enumitem}
\begin{document}

\title{QueryER: A Framework for Fast Analysis-Aware Deduplication over Dirty Data}

\author{Giorgos Alexiou}
\email{galexiou@athenarc.gr}
\affiliation{%
  \institution{IMSI, ATHENA Research Center}
  \institution{School of Electrical and Computer Engineering, NTUA}
  \country{}
}

\author{George Papastefanatos}
\email{gpapas@athenarc.gr}
\affiliation{%
  \institution{IMSI, ATHENA Research Center}
  \country{}
}

\author{Vasilis Stamatopoulos}
\email{bstam@athenarc.gr}
\affiliation{%
  \institution{IMSI, ATHENA Research Center}
  \country{}
}

\author{Georgia Koutrika}
\email{georgia@athenarc.gr}
\affiliation{%
 \institution{IMSI, ATHENA Research Center}
 \country{}
 }

\author{Nectarios Koziris}
\email{nkoziris@cslab.ece.ntua.gr}
\affiliation{%
  \institution{School of Electrical and Computer Engineering, NTUA}
  \country{}
  }

\renewcommand{\shortauthors}{Trovato and Tobin, et al.}

\begin{abstract}
 \silence{Entity Resolution (ER) constitutes a fundamental task for data integration which aims at matching different representations of entities coming from various sources. Due to its quadratic complexity, it typically scales to large datasets through approximate methods, i.e., blocking. In traditional settings, it is a pre-processing step prior to making “dirty” data, available to analysis. With the increasing demand of real-time analytical applications, recent research considers new approaches for integrating Entity Resolution with Query Processing. }In this work, we explore the problem of correctly and efficiently answering complex SPJ queries issued directly on top of dirty data. We introduce \framework, a framework that seamlessly integrates Entity Resolution into Query Processing. \framework executes analysis-aware deduplication by weaving ER operators into the query plan. The experimental evaluation of our approach exhibits that it adapts to the workload and scales on both real and synthetic datasets.
\end{abstract}

\maketitle
\pagestyle{empty} % removes running headers

\section{Introduction} \label{Introduction}

\silence{Exploration and analysis of dirty data have gained great attention recently due to the emergence of data aggregators; i.e., organizations that harvest, aggregate and analyze data containing overlapping and usually contradicting information from multiple sources .}

Analysis-aware data processing refers to an exploratory analysis scenario, where users apply  traditional data integration methods, such as cleaning\cite{Wang14,Giannakopoulou@sigmod20}, during query time. Several approaches extend the capabilities of SQL engines with operators that \textit{relax} the results of the query, by repairing inconsistent data\cite{Giannakopoulou@sigmod20}. \textit{Analysis-aware Entity Resolution} (ER)  is a special case which aims at extending the results of the query by resolving duplicate entities (records that represent the same real-world entity) during query time \cite{Altwaijry13, Altwaijry15, Alexiou19}.
In traditional data integration settings, ER techniques are employed in a pre-processing step, attempting to clean the entire dataset (batch process) before data becomes available for analysis. Such approaches, however, are often inexpedient for many modern \textit{analysis-aware}  applications that need to minimize the \textit{time-to-analysis} by processing only a subset of the entire dataset and produce quick results. For instance, such applications include data aggregators and data virtualization environments, which aggregate and analyze data containing heterogeneous and usually overlapping/contradicting information from multiple sources. 

State of the art in this area attempts to embed ER functionality (e.g., blocking, entity matching) in the query execution pipeline; however, they either lack performance as they apply ER on the results after query execution \cite{Alexiou19}, they offer limited analysis capabilities addressing simple SP queries on single entity collections \cite{Altwaijry13, Alexiou19}, or they require pre-processing steps for transforming the data for more complex analysis (e.g., SPJ queries) \cite{Altwaijry15}. The main challenge we aim to address in this paper, involves the extension of current SQL engines with ER functionality for the analyst to perform  exploratory analysis over multiple sources (i.e., SPJ queries) with no preparation overhead (e.g., data wrangling, clustering etc.). The technical problems we try to answer in this respect are: i) \textit{How do we implement ER techniques from traditional batch processing settings as relational operators that can be executed in query evaluation pipelines}; ii) \textit{What are the new semantics an SPJ query needs to support for fetching and resolving duplicates during evaluation}, and finally iii) \textit{What is the cost of ER operators and how do we consider it in the relational query planning and optimization}.

This work attempts to address the aforementioned problems by introducing \framework, a framework that integrates ER operations into the planning and execution of SPJ queries. To achieve that, we propose \textit{three novel (ER-specific) query operators}, which (\emph{a}) identify and resolve duplicates within a table by employing a schema-agnostic resolution approach with no configuration overhead; (\emph{b}) join duplicate entities between two or more tables and (\emph{c}) group/merge deduplicated entities into a single representation. We then provide a \textit{method for integrating these operators into query execution}. 
For the ER part, the operators employ Blocking and Meta-Blocking techniques to resolve the duplicates while reducing the cleaning overhead. These techniques are traditionally employed in an end-to-end offline setting and are proven to achieve high levels of recall  \cite{Papadakis13, Alexiou15}. To our best knowledge, this is the first work that considers the integration of these techniques into an online setting.
Next, considering that our preliminary experiments confirmed (see Table \ref{tbl:time_breakdown}) that the cost which dominates query execution is the cost of the pairwise comparisons between entities, we propose a \textit{cost-based planner}, which aims at minimizing \textit{the number of comparisons} among alternative query plans. We have implemented our concepts into an SQL query engine, which can be either integrated in any modern relational RDBMS or directly used over raw data files (e.g. csv). We have evaluated our techniques over real and synthetic data measuring the performance and the scalability of our approach. 

In brief, the main contributions of this paper include: 

%users to perform 
\squishlist
\item A novel framework, called \textbf{\framework}, that enables analysis-aware entity resolution over duplicate data with \textit{minimum pre-processing time} and \textit{no configuration overhead}. 
\item Three novel \textit{operators} that implement \textit{core ER operations} (Blocking/ Meta-Blocking, resolution and grouping of results) within a query plan pipeline.
\item A \textit{cost-based planner} for the efficient execution of joins over duplicate data.
\item A thorough experimental evaluation of the proposed solution over real and synthetic datasets.
\squishend 

\textbf{Outline}. Section \ref{Motivation} presents the motivating example used in this paper, Section \ref{Overview} provides an overview of our approach and Section \ref{Preliminaries} presents the basic concepts. Section \ref{Problem Statement} formulates the problem. Section \ref{Operators} introduces the new operators, and Section \ref{ER Query Evaluation} describes the query evaluation methods. Section \ref{Related Work} discusses related work, and Section \ref{Experimental Evaluation} provides the experimental evaluation. Section \ref{Conclusions} concludes the paper.

\begin{table*}[!htb]
\centering
\footnotesize
\begin{tabular}{lllll}
 \toprule
Id & Title                                                    & Author                  & Venue                 & Year \\
\midrule
$P_1$  & 
Collective Entity Resolution  &            & EDBT                  & 2008 \\
$P_2$  & Collective E.R.                                   & Allan Blake             & International Conference on Extending Database Technology & 2008     \\
$P_3$  & Entity Resolution on Big Data                            & Jane Davids, John Doe  & ACM Sigmod            & 2017 \\
$P_4$  & E.R on Big Data                                          & J. Davids, J. Doe       & Sigmod     &      \\
$P_5$  & Entity Resolution on Big Data                            & J. Davids, John Doe.      & Proc of ACM SIGMOD    & 2017 \\
$P_6$  & E.R for consumer data                     & Allan Blake, Lisa Davidson & EDBT                  & 2015 \\
$P_7$  & Entity-Resolution for consumer data                                  & A. Blake, L. Davidson      & International Conference on Extending Database Technology & \\
$P_8$ & Entity-Resolution for consumer data &	Allan Blake , Davidson Lisa &	EDBT &	2015 \\

\bottomrule
\end{tabular}
\caption{Data Table Publications \textit{P}}
\label{tbl:1} 
\end{table*}

\begin{table}[]
\begin{minipage}{\columnwidth}
\centering
\tiny
\begin{tabular}{llllll}
 \toprule
Id & Title                                          & Description                                       & Rank & Frequency & Est. \\
\midrule

$V_1$  & \vtop{\hbox{\strut International Conference on} \hbox{\strut Extending Database Technology}}                & Extending Database Technology & 1    & annual    & 1984       \\
$V_2$  & SIGMOD                                         & ACM SIGMOD Conference                             & 1    &           & 1975       \\
$V_3$  & ACM SIGMOD                                     &                                                   & 1    & annual    & 1975       \\
$V_4$  & EDBT                                            & \vtop{\hbox{\strut International Conference on} \hbox{\strut Extending Database Technology}}        &      & yearly    &        \\
$V_5$  & CIDR                                           & \vtop{\hbox{\strut Conference on  Innovative} \hbox{\strut Data Systems Research}}   &      & biennial  & 2002       \\
$V_6$  & \vtop{\hbox{\strut Conference on  Innovative} \hbox{\strut Data Systems Research}} &                                                   & 2    & biyearly  & 2002    \\
\bottomrule
\end{tabular}

\caption{Data Table Venues \textit{V}}
\label{tbl:2} 
\end{minipage}

\begin{minipage}{\columnwidth}
\centering
\tiny
\begin{tabular}{llllll}

 \toprule
Title  & Year & Rank \\
\midrule
Collective Entity Resolution | Collective E.R & 2008 & 1 \\
E.R for consumer data | Entity-Resolution for consumer data & 2015 & 1 \\
\bottomrule
\end{tabular}
\caption{Result of the sample user query based on our approach }
\label{tbl:3} 
\end{minipage}

\end{table}
\section{Motivating Example} \label{Motivation}
To motivate our work, let us consider a data scientist who works for a scholarly data aggregator such as the Open Academic graph\footnote{\url{https://www.microsoft.com/en-us/research/project/open-academic-graph/}} or Openaire\footnote{Openaire  \url{https://www.openaire.eu}} and performs various types of analysis, e.g., impact assessment and citation analysis. The data aggregator harvests, aggregates and analyzes data from various publishers, open archives and data repositories. The data are then mapped to a common schema and aggregated into data files based on their type (e.g., publications, venues, etc.). The same records may be listed in multiple repositories, thus collected data may contain duplicate entries. As aggregation from sources is performed at arbitrary time intervals, a requirement is that the time-to-analysis must be kept low, and thus, no ETL actions, such as batch deduplication, are applied on the data, every time a new source is harvested. Hence, the user wants to be able to perform on-the-fly queries over the dirty data requiring that duplicate entries must be resolved in the results.

\begin{figure}[h]
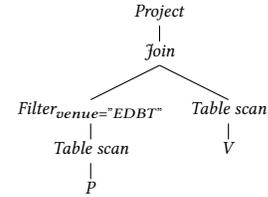

\hspace*{-2cm}                               
\footnotesize
    \Tree[.{\textit{Project}}
        [.{\textit{Join}}  
            [.{ \textit{Filter_{venue="EDBT"}}} 
                [.{ \textit{Table scan} } 
                    [.{ \textit{P} } ]
                ]
            ]
            [.{\textit{Table scan}}
                [.{ \textit{V} } ]
            ]
        ]
    ]
    \caption{Query Plan of the motivating example }
    \label{diag:qplan}
\end{figure}

A part of the collected information concerning publications $P$ and venues $V$ is shown in Tables \ref{tbl:1} and \ref{tbl:2}.
[$P_1$, $P_2$], [$P_3$, $P_4$, $P_5$] and [$P_6$, $P_7$, $P_8$] are sets of the matching publications, coming however from different sources and thus exhibiting differences in the values of their attributes, e.g., author or venue names are abbreviated, some entities have missing years, etc. Respectively, [$V_1$, $V_4$], [$V_2$, $V_3$] and [$V_5$, $V_6$] are sets of matching venues. The user would like to find the publications published in conferences along with the venue rank. She is not aware of the way each data source describes the title of the venue; thus she starts by a query of the form:\newline 
\begin{small}
\texttt{SELECT P.Title, P.Year, V.Rank \newline 
FROM P INNER JOIN V ON P.venue = V.title \newline 
WHERE P.venue="EDBT"}.
\end{small}
\newline 
The execution plan of the user's query is depicted in Fig.\ref{diag:qplan}. The query will first perform a table scan in $P$, select [$P_1$, $P_6$, $P_8$], retrieve $V_4$ via the join with $V$, and finally will provide as output the projected attributes of [$P_1$-$V_4$], [$P_6$-$V_4$] and [$P_8$-$V_4$] joins. However, the query will not include entities $P_2$ (duplicate of $P_1$), $P_7$ (duplicate of $P_6$, $P_8$), and $V_1$ (duplicate of $V_4$), meaning that the user will miss the \textit{Title} and \textit{Year} for $P_2$, $P_7$, and the \textit{Rank} for $V_4$. To capture the requirements of her analysis, the user would expect to view the result of Table \ref{tbl:3}, where duplicate entities have been identified and grouped into a single record;  e.g., values being the same across records are grouped, missing values (null) are replaced by existing ones and contradicting ones are all presented to the user. This is an indicative way of grouping the records and other ways can also be applicable.
In order to view the above results, the user would have to employ traditional (batch) \textit{ER} techniques to fully deduplicate both tables prior to querying them. However, deduplicating the entire collections of $P$ and $V$, after each harvesting process, is unnecessary and time-consuming since only a small subset of the data corresponds to the records that affect the user’s query.\par
An ideal solution for this predicament would (a) identify the records that are affected by the user's query (b) deduplicate them against other records in the database, and (c) join and present the results to the user in a meaningful way. However, such operations are not inherently supported by the conventional SQL operators. To tackle this problem, we have to integrate \textit{Entity Resolution} into \textit{Query Processing} by introducing relational operators that leverage traditional ER techniques into the SQL's query plan pipeline. This task is not straightforward since we have to ensure that (a) for a given query we produce the exact same results as in the case the user performs queries over a database, which has been resolved via a batch ER approach and (b) the overall execution-time is always better compared to the batch ER approach.\par
To address these challenges, we propose \framework, a  framework  that  integrates  ER operations into the planning and execution of SPJ queries. With \framework, the user, 
instead  of  fully  cleaning  the  data  before issuing the query, will issue the exact same query directly on top  of  the  dirty  data,  without  the  need  of  a  pre-processing step  (e.g.  ETL,  batch  deduplication  etc.).  Our  framework will answer  the  user’s  query  and  produce  the  same results  with  the  batch  ER  approach  and  reduce  the  cleaning overhead  by  deduplicating  only  those  parts  of  the  data  that influence  the  query’s  answer. Furthermore, to reduce the cleaning overhead, the cost-based planner will decide the  best  operators  placement,  based  on  ER-related  statistics, with the sole purpose of reducing the total number of pair-wise comparisons that a query will perform.
\silence{ An overview of the proposed framework is depicted in Fig.\ref{fig:evaluation} and the internals of the \textit{Query Executor} in Fig.\ref{fig:dedupExample}.}

\section{Query Evaluation Overview} \label{Overview}
\begin{figure}[h!]
\includegraphics[scale=0.25]{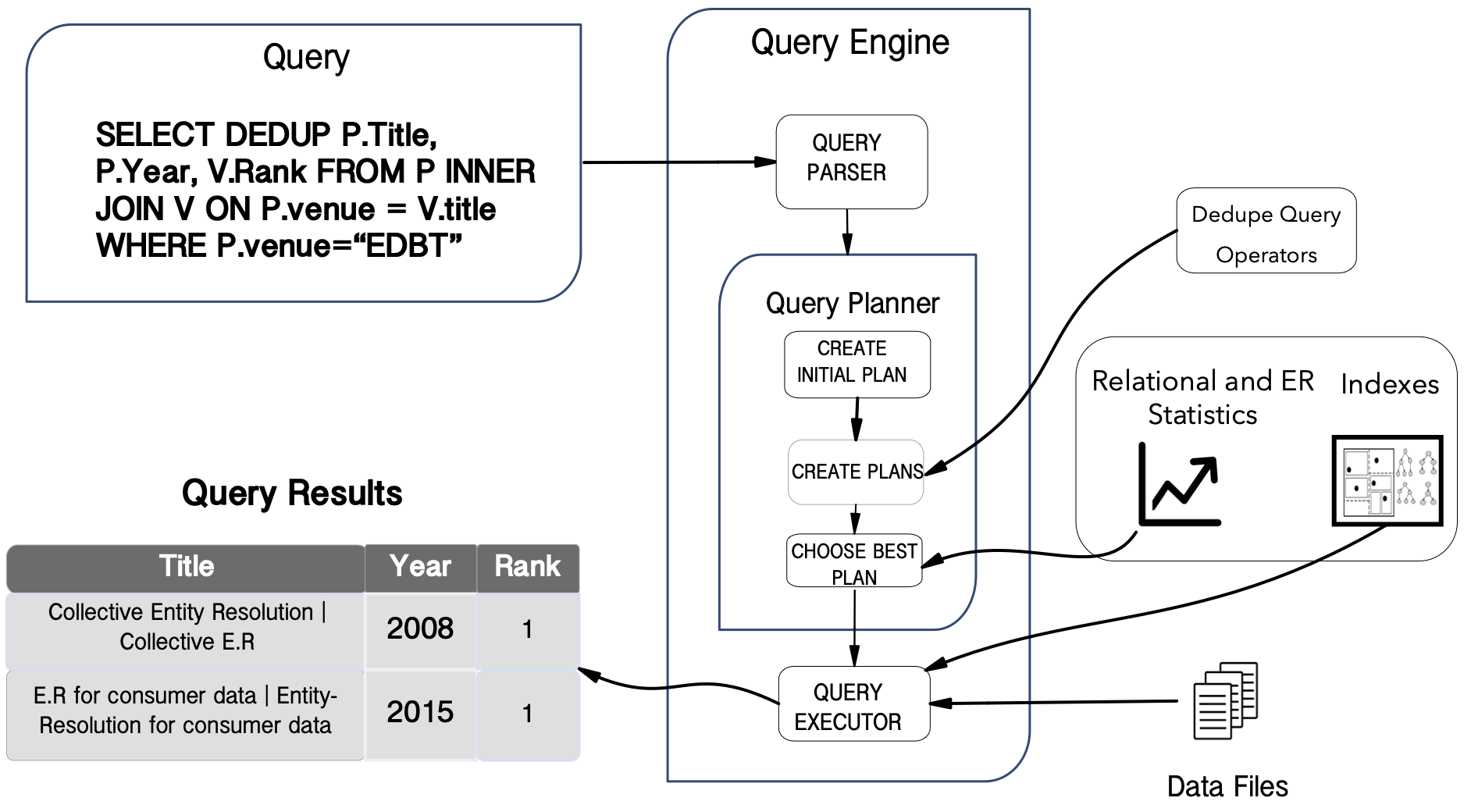}
\caption{Query Evaluation Overview}
\label{fig:evaluation}
\end{figure}

In this section, we provide an overview of our introduced query evaluation flow, which is depicted in Fig. 
\ref{fig:evaluation}\silence{\RED{and the internals of the \textit{Query Executor} in Fig. 
\ref{fig:dedupExample}  - [GP], Why do you refer to FIG4, since you do not explain it. Remove the reference or 
state that this is explained in section XX}}.\par A user utilizes the SQL syntax for querying one or more 
tables with duplicates. The use of the \texttt{DEDUP} keyword in the beginning of the \texttt{SELECT} clause 
denotes that the results should be resolved for duplicates before being returned to the user; otherwise the 
typical SQL semantics are used. The input query is first parsed by the \textit{Query Parser}. It produces an 
abstract representation of the query, which is then used by the planner to create alternative query plans. The 
\textit{Query Planner} creates an initial plan without considering the new \textit{ER operators}. Consequently, this plan is transformed with insertions and substitutions of the \textit{ER operators} to a set of alternative query plans. The planner combines pre-computed relational and ER-specific statistics (e.g. selectivity, 
estimated number of comparisons, etc.) for devising the best plan. It, then, passes this plan to the 
\textit{Query Executor} for the retrieval of the data, the execution of the query operators and the 
presentation of the deduplicated results to the user. For the execution of the operators, \framework also 
utilizes three light-weight indices per table. The \textit{Table Block Index} denoted as $TBI_E$ is a hash 
index that maps a block to a set of record ids.  The \textit{Inverse Table Block Index}, denoted as $ITBI_E$, 
is a hash index that maps record ids to blocks. These indices are sorted in ascending order by their block 
size. The \textit{Link Index} denoted as $LI_E$ stores the link-sets of each entity. All indexes are built 
once-off during initialization of each table and are stored in memory. More details are provided in Section 
\ref{ER Query Evaluation}.  
\silence{
\RED{[GP] NOTE that for the following paragraph: Notation has not yet been introduced but you refer to it as known.E.g., entity collection, entity ids (is it diff from record ID), blocking, DEDUPE query. Simply refer to it in known terms, table, recordID, index and afterwards in Preliminaries you put the notation. I moved everything in a single paragraph} \textit{Blocking and Indexing}. This first step creates two indices per entity collection $E$, which are used in query processing. The \textbf{Table Block Index} denoted as $TBI_E$, which is a hash index that maps a block to a set of entity ids and constructed via a blocking function by all entities in $E$. The \textbf{Inverse Table Block Index}, denoted as $ITBI_E$, is also a hash index that maps entities to blocks. Both indices are sorted in ascending order by the size of its blocks and are built once-off during the initialization of each table and are loaded in-memory during query execution.\par
\textit{Query Planning}. When a user issues an SQL query, the query engine creates an initial plan without considering the ER process. Consequently, the initial plan is transformed with \textit{operators} insertions and substitutions to what is considered a Dedupe query plan, in a two step approach. In the first step we collect precomputed statistics (e.g. selectivity, estimated number of comparisons, estimated number of joins, etc.). In the second step the planner utilizes these statistics to devise the best plan.\par
\textit{Query Plan Execution}. Finally, the query engine executes the plan utilizing the indices and the \textit{operators} to produce the final output.}
\silence{$\mathcal{DR}_G$.}
\section{Preliminaries} \label{Preliminaries}
In this section, we provide the preliminary concepts of our approach. The notations are included in Table \ref{notations}.

\textbf{Entity Collections and duplicate entities.}
Let $D$ be a set of entity collections, i.e., $D$ = $\{E_1,E_2,...,E_n\}$. Each collection $E$ is described by a list of attributes $A^E$ = $\{a^E_1, a^E_2, ..., a^E_k\}$, and contains a set of entities $e_i$, i.e., $E$= $\{e_1$, $e_2$, ..., $e_{|E|}\}$. An entity $e$ is uniquely identified by one of its attributes, denoted as $e_{id} \in A^E$. An entity collection can be a raw data file (e.g. a csv, parquet) or a relational table, although no PKs and FKs are considered in this work.

Two entities $e_{i}$, $e_{j} \in E$  are considered to be \textit{duplicates},  notated with $e_i\equiv e_j$ if they represent the same real-world object. A collection $E$ is called \textit{dirty} if it contains at least a pair of duplicate records.\silence{, i.e., $\exists e_i,e_j\in E, i\neq j:e_i\equiv e_j$.} %and \textit{deduplicated} otherwise.
An example of a dirty entity collection is the \textit{Publications} collection of Table \ref{tbl:1}, where entities $e_{P_1} \equiv e_{P_2}$.

\textbf{Entity Resolution (ER)}. ER is the task of \textit{identifying} and \textit{linking} different manifestations of the same real-world object \cite{Getoor12}. This task is called Deduplication\cite{Pchris12, Pchris122} in our context of homogeneous entity collections. Formally, given a dirty collection $E$, ER is a function that operates on the entities in $E$ and produces a collection of matches (or linkset), denoted as $L_E$, representing the pairs of duplicates in $E$\silence{i.e., ER: $\mathcal{F}$($E$) \rightarrow $L_E$, and \forall $e_i,e_j\in E, e_i\equiv e_j$: ($e_i,e_j$)\in $L_E$}.\silence{ An entity collection $E$ is called \textit{deduplicated} after an ER function has been applied to it and produced an $L_E$.}

\begin{table}[t]
\centering
\label{tab:notation}
\footnotesize
\setlength\extrarowheight{0.99mm}
\begin{tabular}{||l|l||}
% \tline
\hline
\textbf{Symbol} & \textbf{Description}\\ %\dline
\hline
$E$, $e_i$ & An Entity collection,  single entity\\
% $E_G$ & A deduplicated grouped Entity Collection\\
$B$, $b_i$, & A Block collection, a single  block\\
$|b_i|$, $||b_i||$ & Size (\#entities) and cardinality (\#comparisons) of $b_i$\\
% $|B|$, $||B||$ & Size (\#blocks) and cardinality (\#comparisons) of $B$\\
% $Q$  & A flat SQL query\\
% $R_G$ & Deduplicated grouped entities retrieved by a $Q$\\
$\mathcal{DQ}$  & A Dedupe query\\
${QE}_E$, $|{QE}_E|$ & Entities evaluated by a $\mathcal{DQ}$, Size (\#entities) of ${QE}_E$\\
$\overline{{QE}}_E$ & Duplicate Entities of ${QE}_E$\\
$L_E$ & A collection of matching pairs ($e_i$,$e_j$) in $E$ \\
$\mathcal{DR}_E$ & Contains $QE_E \cup \overline{QE}_E$ and $L_E$\\ 
$\mathcal{DR}_G$, & Deduplicated grouped entities retrieved by a $\mathcal{DQ}$\\
$TBI_E$ &  Table Block index for entity collection $E$ \\
%  $ITBI_E$ & Inverse Table Block index for $E$\\
 $QBI_{Q_{E}}$ & A Query Block Index \\
%  $EQBI_{Q_{E}}$ & An enhanced Query Block Index \\
 $LI_E$ & A Link Index for E \\
\hline
\end{tabular}
\caption{Notation for Concepts}
\label{notations}
 \end{table}

\textbf{Blocking.} Blocking is widely used to scale ER \cite{Pchris122,Baxter03} by restricting the executed comparisons to similar entities. The basic concept is the block $b$ = ($e_1,e_2,... e_{|b|}$), which is identified by a unique key ($BK$) and groups entities based on the similarity/equality of their keys (e.g., tokens, n-grams etc.). This way, ER performs pair-wise comparisons only between the entities in $b$, rather than between all entities in $E$.
A set of blocks is called block collection $B$. Its size $|B|$ denotes the number of blocks it contains, while its cardinality denotes the total number of comparisons it involves: $||B||$ = $\Sigma_{bi\in B}||b_{i}||$, where $||b_{i}||$ is the cardinality of a block. 
\silence{Blocking methods rely on redundancy; a blocking key $BK$ places multiple entities in the same block, which results in many redundant (i.e., non-matching) and superfluous (i.e., existing in multiple blocks) comparisons. This way high recall is achieved at the cost of more comparisons (i.e. lower precision). This effect is partially improved by meta-blocking techniques that discard either entire blocks or specific comparisons, within the blocks, during the resolution process.}

\textbf{Meta-Blocking.} Meta-blocking aims at restructuring a block collection $B$ into a new one that contains fewer redundant and non-matching comparisons, while keeping the original number of matching ones \cite{Papadakis133}.\silence{The quality of a block collection $B$ is measured in terms of two competing criteria: \textit{efficiency} and \textit{effectiveness}. The former is directly related to the cardinality $||B||$.The effectiveness of $B$ depends on the cardinality of the set $L_E$ of detected matches (i.e., pairs of duplicate entities compared in at least one block). There is a clear trade-off between the effectiveness and the efficiency of $B$: the more comparisons are executed (i.e., higher $||B||$), the higher its effectiveness gets (i.e., higher $|L_E|$), but the lower its efficiency is, and vice versa. Successful block collections achieve a good balance between these two competing objectives.}
Block processing methods are divided into \silence{according to the granularity of their functionality}\cite{Papadakis13}: i) \textit{Block-refinement},\silence{, which operate at the coarse level of individual blocks,} and ii) \textit{Comparison-refinement} methods\silence{, which operate at the finer level of individual comparisons}.

From the former category, we employ \textit{Block Purging (BP)} and \textit{Block Filtering (BF)} \cite{Papadakis162}. Both methods rely on the idea that the larger a block is, the less likely it is to contain unique duplicates.
 $BP$ aims at cleaning the block processing list from oversized blocks \silence{which involve a large number of unnecessary comparisons. These are blocks }that correspond to tokens of little discriminativeness, thus entailing a large number of unnecessary comparisons. \silence{of comparisons while being unlikely to contribute non-redundant matches.Hence, they can be safely excluded from the block processing step, enhancing considerably the efficiency without any significant impact on the effectiveness.} 
 For instance, consider the token "Entity" in Table \ref{tbl:1}: it contains  most of the possible comparisons of the entities and the only non-redundant comparisons it involves are non-matching. The $BF$ method, unlike $BP$ though, is applied independently to the blocks of every entity, assuming that each block has a different importance for every entity it contains. Based on this idea, $BF$ restructures a block collection $B$ by removing entities from blocks, in which their presence is not required.\silence{The importance of a block for an individual entity $e_i$$ \in $$b_i$ is determined by the maximum number of blocks $e_i$ participates in (the fewer the better).$BF$ requires a \textit{filtering ratio r} parameter, which determines the portion of the most important blocks that are retained for each entity and it is defined in the interval [0, 1] (e.g. r=0.5 means that each entity remains in the first half of its associated blocks, after sorting them in ascending cardinality).}\par
From the later category, we employ the \textit{Edge Pruning (EP)} method \cite{Papadakis162}. This method restructures a block collection $B$ into a new one that contains significantly fewer unnecessary comparisons, while maintaining almost the same effectiveness. It operates in two steps: (i) it transforms $B$ into a blocking graph, which contains a node \silence{$v_i$}for every entity\silence{$e_i$$\in$$B$} and an edge for every pair\silence{ ($e_i$, $e_j$)} of co-occurring entities,\silence{ The same pair of entities in multiple blocks is mapped to a single edge,} (ii) it annotates every edge with a weight analogous to the likelihood that the incident entities are matching. Therefore, $EP$ discards most superfluous comparisons by pruning the edges with low weights.\silence{ For the weighting scheme we are based on the fundamental property that two entities are more likely to match, when they have many blocks in common.}

\textbf{Entity Matching and Grouping.} Following the best practices in the literature, we consider entity matching as an orthogonal task to blocking \cite{Pchris12,Pchris122,Papadakis13,Alexiou15}. That is, we assume that two duplicates $e_i \equiv e_j$ are detected in $B$ as long as they co-occur in at least one of its blocks. As the vast majority of duplicate entities co-occur, the actual performance of ER depends on the accuracy of the similarity method used for entity comparison (e.g., Jaccard, Jaro-Winkler, etc). 

The final step of an ER task involves the grouping of duplicate entities into a single representation. Given a dirty collection $E$ and a linkset $L_E$, a grouping function\silence{$\matchcal{G}$($E$,$L_E$)\rightarrow $E_G$} produces a set of \textit{deduplicated grouped entities} $E_G$. Several fusion techniques can be applied, such as grouping duplicates based on a surrogate key or fusing the values of the entities' attributes based on domain-specific rules\cite{Dong2014FromDF}. We consider this problem as an orthogonal task, since the performance of grouping depends on the fusion technique employed.\silence{Next, we provide the basic notion of our approach and formulate the problem we try to solve.}

\section{Problem Statement} \label{Problem Statement}
In this section, we present the semantics of the query answering by introducing the notion of \textit{Dedupe Query} and formulate the  problem. 

\textbf{Dedupe Query}. Let $Q$ be a flat SQL query over a set of \textit{dirty} entity collections 
$E_1$,$E_2$,...$E_n$ in $D$. We consider conjunctive and disjunctive queries where a condition expression can 
be of the form: $E.x$ \texttt{op} \texttt{constant} ($op$ can be $=$,$>$,$<$, IN, etc) or $E_1.x = E_2.y$ 
(equijoins). Let ${QE}_{E_1}$,${QE}_{E_2}$,..., ${QE}_{E_n}$ be the sets of entities evaluated by $Q$ after all conditions, from the \texttt{WHERE} clause, are applied to $E_1$,$E_2$,...$E_n$. \silence{respectively and 
$QE_E$ = $QE_{E_1}$\bowtie $QE_{E_2}$\bowtie ...\bowtie $QE_{E_n}$ be the final set of joined entities 
evaluated by $Q$.}Let, also, $\overline{QE}_{E_1} \subseteq E_1$, $\overline{QE}_{E_2}\subseteq E_2$, ..., 
$\overline{QE}_{E_n} \subseteq E_n$ be the sets of entities which are not evaluated by $Q$ but have duplicates 
in ${QE}_{E_1}$,...,${QE}_{E_n}$, respectively, i.e., $\forall e_j \in$ $\overline{QE}_{E_k}$: $\exists$ $e_i\in {QE}_{E_k}$: $e_i\equiv e_j$. %We denote the extended set of matching entities as $Q^{l}_{E_k}$ = $Q_{E_k}$$\cup$$\overline{Q}_{E_k}$.

Then, a \textsf{Dedupe Query} $\mathcal{DQ}$ is the \textit{deduplication query}, equivalent to $Q$, over $D$, which operates on the different sets ${QE}_{E_1}$,${QE}_{E_2}$,..., ${QE}_{E_n}$, produces the deduplicated sets $\mathcal{DR}_{E_1}$,$\mathcal{DR}_{E_2}$,..., $\mathcal{DR}_{E_n}$, with $\mathcal{DR}_{E_k}$ = $<$${QE}_{E_k}\cup\overline{{QE}}_{E_k}, L_{E_k}$$>$, $k = 1..n$ and returns a set of \textit{deduplicated grouped entities}, denoted as $\mathcal{DR}_G$. %  which are the grouped duplicates of the entities into single records.

\silence{Specifically, for the set of entities ${QE}_{E_k}$ that are evaluated over ${E_k}$ , $\mathcal{DQ}$ applies the ER function\silence{: $\mathcal{F}$(${QE}_{E_K}$) \rightarrow $L_{E_k}$, in order } to find the matching pairs $L_{E_k}$, identify the duplicates \overline{QE}$_{E_k}$ and create the union of the two entity sets.\silence{Next, it performs the joins over the deduplicated sets to produce a set of joined entities $\mathcal{DR}_E$ = $\mathcal{DR}_{E_1}$\bowtie $\mathcal{DR}_{E_2}$\bowtie ...\bowtie $\mathcal{DR}_{E_n}$.} Finally, it  applies a grouping function\silence{ $\matchcal{G}$($\mathcal{DR}_E$, $L_{E_1}$,...,$L_{E_n}$)} to produce the final set of deduplicated \textit{grouped} entities $\mathcal{DR}_G$ and projects the attributes that are included in the \texttt{SELECT} clause.}
 
\textbf{Batch Approach Query (BAQ)}. 
Given a set of entity collections $D'$=$\{E_{G_k}\}$, produced offline via a batch ER operation on $D$=$\{E_k\}$ (Sec. \ref{Preliminaries}). A $\mathcal{BAQ}$ is an SQL query which operates on the \textit{set of deduplicated grouped entities $E_G$} and returns the result-set $R_G$.\silence{ the $BA$ operates in an offline/batch mode and applies the exact same ER and grouping functions with $\mathcal{DQ}$, on the different entity collections $E_1$,$E_2$,...$E_k$ in $D$. The $BA$ deduplicates the entire collections offline and produces a set of deduplicated \textit{grouped} entity collections $D'$=$\{E_{G_k}\}$. Every query $\mathcal{BAQ}$ issued on $D'$, returns a subset of these deduplicated \textit{grouped} entities $R_G$.}

\textbf{Problem Statetement}. Given an $R_G$ and a $\mathcal{DR}_G$ being the returned deduplicated \textit{grouped} entities for $\mathcal{BAQ}$ and $\mathcal{DQ}$ respectively. Then, our setting is a query optimization problem s.t.
\silence{Given a $\mathcal{DQ}$ issued over a set of \textit{dirty} entity collections $D$=$\{E_k\}$ and a $Q$ (equivalent to $\mathcal{DQ}$) issued over a set of deduplicated \textit{grouped} entity collections $D'$=$\{E_{G_k}\}$ (produced via a batch ER approach from $D$). With, $R_G$ and $\mathcal{DR}_G$ being the returned deduplicated \textit{grouped} entities for $Q$ and $\mathcal{DQ}$ respectively. Then, our setting is a query optimization problem s.t.} 
\silence{Let $D$=$\{E_k\}$ be a set of \textit{dirty} entity collections and $D'$=$\{E_{G_k}\}$ be a set of deduplicated \textit{grouped} entity collections. With, $Q$ be a \textit{flat SQL} query issued over a set of deduplicated \textit{grouped} entities $E_G$$\in$$D'$\silence{the set of \textit{deduplicated} entity collections $D'$=$\{<E_k,L_{E_k}>\}$,}, produced via a batch ER task from $D$, and $R_G$ being the returned deduplicated \textit{grouped} entities. Also, let $\mathcal{DQ}$ be the equivalent to $Q$ Dedupe query, applied on the original set of \textit{dirty} collections $D$. Our setting is a query optimization problem s.t.}
\begin{enumerate}[noitemsep,topsep=0pt]
\item ($\mathcal{DQ}$ Performance) The \silence{ER efficiency is maximized and the}execution time of $\mathcal{DQ}$ over $D$ is less than the execution time of $\mathcal{BAQ}$ over $D'$ plus the time needed for applying ER over the entire $D$ and producing $D'$.
\item ($\mathcal{DQ}$ Correctness) The set of entities returned by $\mathcal{DQ}$ over $D$ equals to the set of entities returned by the evaluation of $\mathcal{BAQ}$ over $D'$; i.e., $\mathcal{DR}_G \equiv$ $\mathcal{R_G}$. \end{enumerate}
\section{Dedupe Query Operators}\label{Operators}

In this section, we introduce three novel query operators: (i) Deduplicate, (ii) Deduplicate-Join and  (ii) Group-Entities. These operators are the building blocks of a \textit{Dedupe Query} implementation.% as they implement \textit{Entity Resolution} methods into the SQL’s query execution.

 \subsection{Deduplicate Operator} \label{Deduplicate Operator}
\begin{figure}[h]
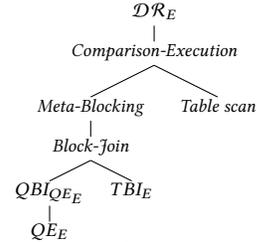

\hspace*{-2cm}                                
\footnotesize
	\Tree
		[.{\textit{$\mathcal{DR}_E$}}
		[.{\textit{Comparison-Execution}}
		[.{ \textit{Meta-Blocking}}
			[.{ \textit{Block-Join}}
				[.{ \textit{$QBI_{QE_E}$} }
				[.{\textit{$QE_E$}}
				]]
				[.{ \textit{$TBI_E$} }
				]
		]]
		[.{\textit{Table scan}}
			]
		]]
    	\caption{A typical Deduplicate Operator pipeline}
	\label{dia:dop}
\end{figure}

\begin{figure*}[h!]
\hspace*{-1cm} 
\includegraphics[scale=0.58]{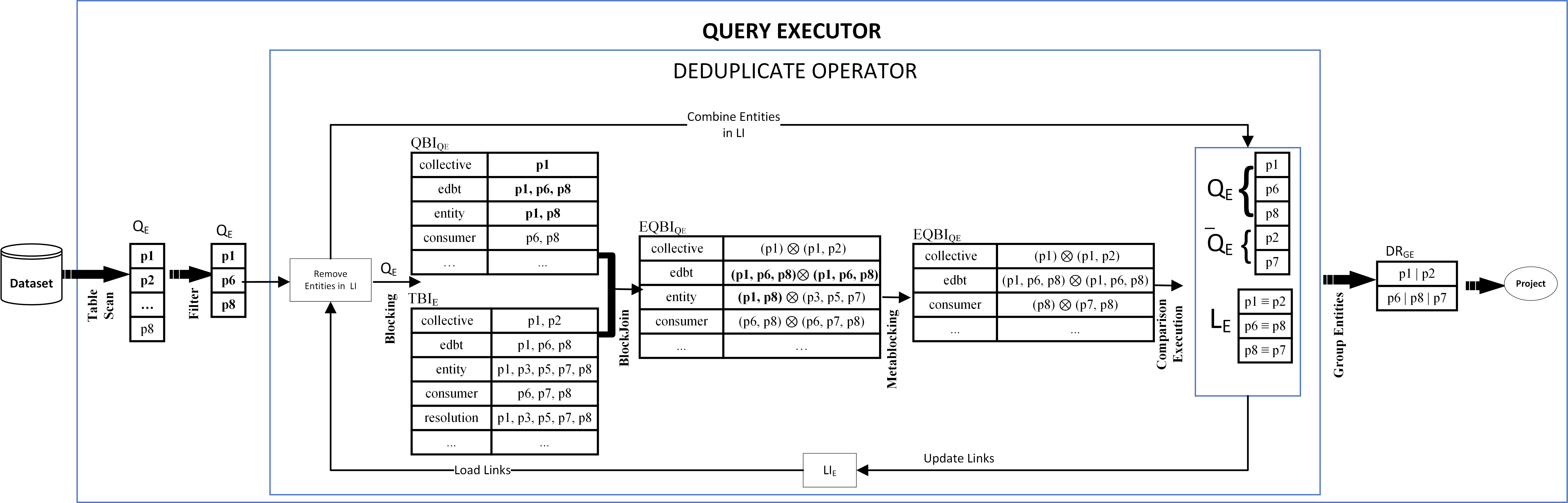}
\caption{Description of the Deduplicate Operator including Block-Join, Meta-Blocking and Comparison-Execution steps.}
\label{fig:dedupExample}
\end{figure*}

 The Deduplicate Operator is a relational operator that constitutes the key concept of ER integration into the traditional query processing. It  processes a set of entities $QE$$\in$$E$ derived by the user query, and finds their duplicates in $E$. The operator achieves its goal by encapsulating several distinct operations of an Entity Resolution workflow %\cite{Alexiou19} 
 in its pipeline: (i) Blocking, (ii) Block-Join, (iii) Meta-Blocking, (iv) Comparison-Execution as well as other primal relational operations (e.g. Table Scan). The reason that multiple distinct ER operations are encapsulated into a single operator is that their sequence is strict as dictated in an established ER pipeline \cite{Papadakis133}.
 The input of this relational operator is a set $QE_E$ , while the final output is its super-set $\mathcal{DR}_E$. \silence{, which contains all duplicate entities of $QE_E$, i.e., ${\overline{QE}}_E$ and the links between them, i.e., $L_E$.} 
The operator's pipeline is depicted in Fig. \ref{dia:dop}, while a detailed example of its internals\silence{, as part of the overall query processing pipeline} is shown in Fig. \ref{fig:dedupExample}. Before we describe each operation in the operator's pipeline, we describe the indices employed.

The operator makes use of three in-memory indices for managing the block collections. The Table Block Index $TBI_E$ maps a block to a set of record ids (Sec. \ref{Overview}). The \textbf{Query Block Index} $QBI_E$  is a hash index of the blocks, and is constructed on-the-fly for all entities in $QE_E$. Note that both $QBI_E$ and $TBI_E$ are always constructed via the same blocking function and as $QE_E\subseteq{E}$, then $|QBI_{QE_E}|\leq|TBI_E|$. Finally, the Link Index $LI_E$ is a hash index that maps each entity to its duplicate entities (Sec. \ref{Overview}). It is initially empty and is amended with the links that each query resolves. Since the computation of $L_E$ is costly,  $LI_E$ is crucial to the efficiency of our approach because, in each query, we only need to compute the link-sets of those entities in $QE_E$ that are not already in $LI_E$. Thus, our approach tends to get significantly faster with every query issued over the dirty dataset.

The operations of the Deduplicate Operator are presented next.

\textbf{i) Query Blocking.} This operation takes the entities derived from the user query $QE_E$ that are not in $LI_E$, and creates a $QBI_{QE_E}$ by invoking the same blocking function that was used for the construction of the $TBI_E$. In this work, we use the \textit{Token Blocking} \cite{Papadakis13} which is a schema-agnostic blocking method that builds blocks on the occurrence of each different  token found in the values of all attributes of an entity $e$. The tokens serve as the keys ($BKs$) of the blocks. In our example, applying \textit{Token Blocking} on the \textit{title} of $e_{P_1}$ and $e_{P_2}$ entities will form the blocks: $b_{Collective}$ = \{$e_{P_1}$, $e_{P_2}$\}, $b_{Entity}$ = \{$e_{P_1}$\}, $b_{Resolution}$ = \{$e_{P_1}$\}, and $b_{E.R.}$ = \{$e_{P_2}$\}.

\textbf{ii) Block-Join.} The Block-Join operation \cite{Alexiou19} operates on two block collections, $QBI_{QE_E}$ and $TBI_E$. It performs a \textit{hash-join} between the keys ($BKs$) of the two block collections and enriches the blocks of $QBI$ with the set of entities from $E$ that exist in the blocks of $TBI_E$ and share the same blocking keys. The enriched $QBI_{QE_E}$ is called $EQBI_{QE_E}$. This operation is essential in order to retrieve all those “dirty” (i.e. containing duplicates) subsets of entities ${\overline{QE}}_E$ that approximately (possibly containing false-positives but not the opposite) answer the user’s initial query.\newline
\textbf{iii) Meta-Blocking.} The \textit{Meta-Blocking operation} is applied on  $EQBI_{QE_E}$ to drastically reduce any unnecessary comparisons induced by the Block-Join and thus to achieve better efficiency without significantly affecting the effectiveness. More specifically, we employ, in sequence, the \textit{Block Purging} ($BP$), \textit{Block Filtering} ($BF$) and \textit{Edge Pruning} ($EP$) methods. The sequence of applying the methods is strict as operating on the coarse level of individual blocks involves very low space and time complexity in comparison to operating on the finer level of individual comparisons. Thus, removing the blocks that contain unnecessary comparisons first, enhances the performance of $EP$ as the size of the blocking graph after the application of $BP$ and $BF$ is considerably smaller. Between the \textit{Block-refinement methods}, the $BP$ comes first as $BF$ focuses on individual blocks and $BP$ in the whole block collection.

\textbf{iv) Comparison-Execution.} The final step is the execution of the remaining comparisons. In each block, comparisons survived from Meta-Blocking are performed between the entities of $QE_E$ and $EQBI$ sets. This way, we enable the identification of duplicates only for the initial selection $QE_E$ and not for all possible duplicates within a block. Hence, the comparisons are further reduced. % that belong to the $QE_E$ and resolves the matching ones. For each block in our $QBI$, we take all the entities that derived from the user’s select query |$QE$|and we compare them with the entities its entities |$Qb_i$|  that belong to.% \RED{Thus the number of comparisons for each block will be: ||$Qb_{i}$|| = $|QE \cap EQb_{i}| \cdot (|b_{i} - QE$|) and the total comparisons will be: ||QBI||= $\Sigma_{EQb_{i}\in QBI} ||EQb_{i}|| $  \galexiou{SUM SYMBOL DOES NOT APPEAR FOR SOME REASON}}. 
For the actual comparisons, we follow a schema-agnostic approach and we compare the values of all corresponding  attributes between entity pairs. This approach does not discriminate between the attributes whose values may exhibit a higher likelihood for possible  duplicates (identifiers, code lists, etc.) but it requires \textit{no  configuration} from the user; however, any schema-based alternative can be used.\silence{ To further improve our performance, at the time of the resolution we create and keep a hashed key for every comparison performed (based on the two entity ids) to avoid re-executing it. Those hashes are maintained in a data-structure that offers constant-time access.}
% \begin{table}[h!]
%     \begin{minipage}{0.5\linewidth}
%         \caption{$Q^{l}_{E_k}$ Structure}\label{tab:caption}
%         \centering
%         \footnotesize
%         \begin{hashtable}
%               &   p1 & $\rightarrow$ $[p1, attr_1,...,attr_n]$\\
%               &  p2 & $\rightarrow$ $[p2, attr_1,...,attr_n]$\\
%               &  p3 & $\rightarrow$ $[p3, attr_1,...,attr_n]$\\
%               & p4 & $\rightarrow$ $[p4, attr_1,...,attr_n]$\\
%               & p5 & $\rightarrow$ $[p5, attr_1,...,attr_n]$\\
%               &  p6 & $\rightarrow$ $[p6, attr_1,...,attr_n]$\\
%               &  p7 & $\rightarrow$ $[p7, attr_1,...,attr_n]$\\
%               &  p8 & $\rightarrow$ $[p8, attr_1,...,attr_n]$\\
%         \end{hashtable} 
%     \end{minipage}
%     \begin{minipage}{0.45\linewidth}
%       \caption{$L_e$
%       structure}
%       \footnotesize
%       \label{tab:dimFFT}
%         \centering
%         \begin{hashtable}
%           &  p1 & $\rightarrow$ p2\\
%           &  p2 & $\rightarrow$ p1\\
%           &  p3 & $\rightarrow$ p4 $\rightarrow$ p5\\
%           & p4 & $\rightarrow$ p3  $\rightarrow$ p5 \\
%           & p5 & $\rightarrow$ p3  $\rightarrow$ p4 \\ 
%           & p6 & $\rightarrow$ p7 $\rightarrow$ p8 \\ 
%           & p7 & $\rightarrow$ p6 $\rightarrow$ p8 \\
%           & p8 & $\rightarrow$ p6 $\rightarrow$ p7 \\
%         \end{hashtable} 
%     \end{minipage}
% \end{table}

As can be seen in Fig.\ref{fig:dedupExample}, the operator first applies the \textit{Block-Join} to produce  $EQBI_{QE_E}$.  Meta-Blocking follows which reduces both the size $|EQBI_{QE_E}|$ and the comparisons $||EQBI_{QE_E}||$. Finally, the Comparison-Execution performs the Cartesian product of the comparisons between the two sets, within each block. In this step, no comparison will be performed more than once and an entity will not be compared with itself. The result of this operation is a $\mathcal{DR}_{E}$ \silence{which consists of $QE_{E}\cup {\overline{QE}}_E$ and $L_E$}. The final step amends the $LI_E$ with the new links from $L_E$.

$\mathbf{DQ}$ \textbf{Correctness}.
Consider a $\mathcal{BAQ}$ issued over an $E_{G} \in D'$ that returns an $R_G$ and a Dedupe query 
$\mathcal{DQ}$, equivalent to $\mathcal{BAQ}$, issued over an $E$$\in$$D$, that returns a $\mathcal{DR}_G$ 
(see Sec. \ref{Problem Statement}). Given that, i) Blocking, ii) Meta-Blocking are deterministic functions, 
then we have that: i) $TBI_{DQ} \equiv TBI_{BAQ}$ ii) $EQBI_{DQ} \subseteq TBI_{BAQ}$ respectively. Also, for 
each entity $e\in \mathcal{R}_G$, exists a set of blocks $BS_i$=$\{b_1,...,b_m\}$ where i=1...n, that $e$ 
exists in every $b_m\in BS_i$. Hence, the super-set $\mathcal{BS} \equiv {EQBI}_{DQ}$. Consequntly, since 
comparison-execution and grouping are also deterministic functions, then we have that $\mathcal{DR}_G \equiv$
$\mathcal{R_G}$.

\silence{and $E_G$ is produced also by a deterministic function $\matchcal{G}$($E$,$L_E$)\rightarrow $E_G$
 Then, for each entity $e$$\in$$\mathcal{R}_G$, exists a set of blocks $\mathcal{BS}$=$\{b_1,...,b_n\}$ where $e$ exists in every $b_i$$\in$$\mathcal{BS}$ and $\mathcal{BS}$$\subseteq$$TBI_E$. \silence{from where all links \forall $e_i,e_j$$\in$$\mathcal{S}_b$ and $e_i\equiv e_j$ $\exists$ ($e_i,e_j$)\in $L_E$, derived after the execution of all pair-wise comparisons within each $b$.} \textcolor{red}{The next sentence: something is missing} Given the fact that, Blocking, Meta-Blocking, Comparison-Execution and, Grouping are deterministic in both $\framework$ and \textit{batch ER} approaches and the $QBI_{QE_E}$ is formed with the same blocking function as $TBI_E$. Consequently, for a Dedupe query $\mathcal{DQ}$, equivalent to $\mathcal{BAQ}$, we have the $\mathcal{BS}$$\equiv$$EQBI_{QE_E}$ equivalence; i.e., $\mathcal{DR}_G$ \equiv $\mathcal{R_G}$.}

\textbf{Cost Analysis}. The cost of the Deduplicate operator comprises the I/O cost for reading the data from disk, the cost for building the $QBI_{QE_E}$, the Block-Join, the Meta-Blocking techniques, and finally the comparison execution. Regarding the I/O cost, the total number of entities includes: (i) the initial set $QE_{E}$ evaluated by the query and (ii) the set of entities $e$$\in$$EQBI_{QE_E}$ that will be read from disk after Meta-Blocking.\silence{\RED{We denote the number of duplicates w.r.t. the total number of entities in $E$ as the collection's \textit{duplication} factor, $df$ = ${\overline{QE}}_E$ / (${\overline{QE}}_E$+$QE_E$); and we use this factor to estimate the final number of entities that will be read as $\mathcal{DQ}_E$ = $QE_E$*(1+$df$).} \galexiou{ the RED is wrong. The I/O cost is the entities in EQBI after MB but we cannot calculate that size- We need to reconsider the I/O cost since it's not bounded to the planning}} Finally, $TBI_E$ and $LI_E$ access costs are not considered, since they are kept in memory; as shown in Sec.\ref{Experimental Evaluation}, their sizes remain small even for large entity collections.

The computational cost is split between the different parts of the Deduplicate operator. We focus on the ER-specific operations (relational filter and table scan are computationally inexpensive):

\squishlist
\item
$QBI_{QE_E}$ is created by iterating over all attributes $A_E$ of the $QE_E$ entities; for homogeneous data collections,  the cost of $QBI_{QE_E}$ is mainly determined by $|QE_E|$ and $|A_E|$, i.e., $O$($|QE_E|$$\times$$|A_E|$). 

\item The cost of the Block-Join and the two block-refinement methods is determined by the number of blocks in $TBI_E$ and $QBI_{QE_E}$, respectively. The cost of Block-Join is $3$($|QBI_{QE_E}|$+$|TBI_E|$), since we perform a hash-join, the cost of \textit{BP} is $O$($|EQBI_{QE_E}|$) since the blocks are traversed only once. The cost of \textit{BF} is $O$($|EQBI_{QE}|$$\times$$|b_i|$) since for each block $b_i$$\in$$EQBI_{QE_E}$ we iterate all its entities. 

\item \textit{EP} and comparison-execution operate on the pair of entities contained in each block. For 
estimating the number of comparisons within a block $b_i$$\in$$EQBI_{QE_E}$, we must compare all the entities 
$QE_{b_i}$ from this block, which are the entities that intersect with the entities $QE_E$, with all the other entities in the same block. Since we perform a comparison between two entities \textit{only once} and we avoid comparing an entity with itself, then the comparisons will be $|QE_{b_i}|\times$($|b_{i}|$ - 
($|QE_{b_i}|$+1)/2) and for the entire $EQBI_{QE_E}$ is $\Sigma_{b_{i}\in EQBI_{QE_E}}$ 
$|QE_{b_i}|\times$($|b_i|$ - ($|QE_{b_i}|$+1)/2). This number must be multiplied with the cost of the resolution function (e.g. Jaro-Winkler) to assess the cost of the comparison execution.\silence{ Note that, this number depends on the selectivity of the query (the $|QE_{b_i}|$ with respect to the $|b_i|$); as it gets bigger then $|QE_{b_i}|\approx|b_i|$ and the number of comparisons within a block becomes $|b_i|\choose2$.} In practice, the total number is even smaller since the comparisons belonging to multiple blocks will be executed only once, and we only need to compute the link-sets of those entities in $QE_E$ that are not in $LI_E$. Thus, the dominating cost of the comparison execution tends to get significantly lower with each issued query.
\squishend

\subsection{Deduplicate-Join Operator} \label{Deduplicate-Join Operator}
The Deduplicate-Join operator is analogous to the common relational algebra join operators with one exception: it knows whether the input for each side is dirty data or not and consequently performs the corresponding cleaning operations. This operator can accept as input not only deduplicated data ($\mathcal{DR}_{E}$), but also dirty ones ($QE_{E}$) and operates using the two following types:

\silence{\textbf{Problem Definition of Join.} Given a table Join $\bowtie$ between $T_1$ and $T_2$, let J denote the set of entities that satisfy the join $\bowtie$ when both tables have been cleaned prior to the \textit{Join Operation}. Also, let $J^c$ denote the set of entities returned by the \textit{Deduplicate-Join Operator}. Then, we can formally define the problem of join as follows:
\begin{enumerate}
\item Minimize execution time of Join $\bowtie$
% \item Join satisfaction: $\forall$$e_i$$\bowtie$$e_j$$\in$$J^c$, $e_i$,$e_j$ satisfy Join $\bowtie$ 
\item Join Correctness: $J^c$$\equiv$J                                                             
\end{enumerate}
Based on the definition, we introduce two join types:}

\textbf{(i) Dirty-Right.} Takes as input a set  $QE_{E}$ from the right side, a set  $\mathcal{DR}_{E}$ from the left, the join type, which is derived from the already formed query plan, and the attributes that the join is going to be performed on. The operator (Alg. \ref{alg:DJoinOp}) first performs a \textit{relational join} between the  $L$(eft)$\mathcal{DR}_{E}$$\bowtie$$QE_{E}$ and forms a $QE_{E}'$ for the right dirty entity-set (line 4), by discarding the entities that do not join with the left entity-set $\mathcal{LDR}_E$. It then applies the Deduplicate operator's pipeline (line 5) to produce the $R$(ight)$\mathcal{DR}_E$. Finally, after both sides are resolved, it produces the final output by invoking the Deduplicate-Join operation  to join the two resolved sets (line 11).

\textbf{(ii) Dirty-Left.} The operator works in the same way as the Dirty-Right but for the opposite sides.

The reason that we always apply the deduplicate operations on one branch of the query tree prior to the join execution is to satisfy all the possible join predicates (use all possible variations of an entity's values). Like in our motivating example in Sec.\ref{Motivation}, if we try to join two tables without cleaning at least one first, we might end up with missing entities or even an empty set and thus with an incorrect answer. In this case, the $\mathcal{DQ}$ Correctness is also satisfied since we apply the \textit{Deduplicate Operator} to both branches independently, where its correctness has been already proved in Sec.\ref{Deduplicate Operator}.
Independently from the case and the input that each join type takes, the operator always has a consistent output which has the same structure, i.e. $\mathcal{DR}_{E}$. The case-independent output is essential because multiple join operations might need to be performed in a multi-join query plan. The algorithm of the \textit{Deduplicate-Join Operator}  is shown in Alg.
% \vspace{-10pt}
\ref{alg:DJoinOp}.
\newcommand\mycommfont[1]{\footnotesize\ttfamily\textcolor{black}{#1}}
\SetCommentSty{mycommfont}
\begin{algorithm}[h!]
	\scriptsize
	\caption{Deduplicate-Join Operator}
	\label{alg:DJoinOp}
	\algrenewcommand\algorithmicindent{.8em}%
% 	\tcc{JoinType is based on whether or not the Deduplicate Operator has been already applied on each query leaf}
	\SetKwInOut{Input}{Input}
	\SetKwInOut{Data}{Data}
	\Input{$Left$\\$Right$\\$JoinAttributes$\\$JoinType$}
	\KwResult{Joined $DR_E$}
    \uIf{$JoinType$ is DIRTY-RIGHT}{
		$L\mathcal{DR}_E \gets Left$\\
		${QE}_E \gets Right$\\
        $QE_E' \gets discardRight({QE}_E \bowtie L\mathcal{DR}_E)$
		 \\
		$R\mathcal{DR}_E \gets Deduplicate(\mathcal{Q}_E')$
		\\
		
    }
    \uElseIf{$JoinType$ is DIRTY-LEFT}{
        $R\mathcal{DR}_E \gets Right$\\
		${QE}_E \gets Left$\\
	    $QE_E' \gets discardLeft({QE}_E \bowtie R\mathcal{DR}_E$
		\\
		$L\mathcal{DR}_E \gets Deduplicate(\mathcal{Q}_E')$
        \\
    }
	\Return {DeduplicateJoinOperation($L\mathcal{DR}_E$ , $R\mathcal{DR}_E$, joinAttributes)}
\end{algorithm}
% \vspace{-10pt}

The \textit{Deduplicate-Join Operation} (Alg. \ref{alg:DJoin}) takes as input the deduplicated entity sets $L\mathcal{DR}_{E}$ and $R\mathcal{DR}_{E}$, and the join attributes and returns the join between them, denoted as $J\mathcal{DR}_{E}$. 
%It then creates a lookup table of one of the two tables, that hashes its entities using as keys the value(s) of its join-attribute(s) and returns a \textit{QER}. To distinguish the newly created entities each one of them will be given an id through the jId variable. Algorithm \DJoin describes this operation  the right table\newline%
The algorithm starts (line 4) by iterating the $L\mathcal{DR}_{E}$. It first checks if it has visited an entity. If not, it gets its duplicates from the $L_E$ and marks it as visited (lines 6-7). Then, for each of these duplicates, it finds the entities of $\mathcal{RDR}_E$ that join with them, along with their duplicates (lines 8-12). Having found all similar entities that join from both tables, it performs the Cartesian product of these sets (line 14).
Finally, it returns the new joined
% \vspace{-20pt}
$J\mathcal{DR}_{E}$.
\begin{algorithm}
	\scriptsize
	\caption{Deduplicate-Join Operation}
	\label{alg:DJoin}   
	\SetKwProg{Fn}{Function}{}{}
	\Fn{DeduplicateJoinOperation($L\mathcal{DR}_E$, $R\mathcal{DR}_E$, joinAttributes)}{
		$J\mathcal{DR}_{E}$ \tcp{Joined $\mathcal{DR}_{E}$}
		%$JL_E$ \tcp{Joined $L_e$}
% 		$jId \gets 0$ \tcp{Id of new joined entity}
        $visited = set()$\\
		\For {$e \in L\mathcal{DR}_E$}{\label{forins}
% 			$counter \gets jId$ \\
 			\If{$e \notin visited$}{
				$E_{left} \gets e \cup L\mathcal{DR}_E.L_E.get(e)$ \\
				$visited.addAll$($E_{left}$)\\
				\For {$e_l  \in E_{left}$}{
					$E_{joined} \gets e_l \bowtie R\mathcal{DR}_E$ \\
					\For{$e_r \in E_{joined}$}{
						$E_{right} \gets e_r \cup R\mathcal{DR}_E.L_E.get(e_r)$ \\
					}
				}
				$J\mathcal{DR}_E.add( E_{left} \times E_{right}$) 
				% \For{$e_l  \in E_{left}$}{
				% 	$counter \gets jId$\\
				% 	\For{$e_r \in E_{right}$}{
				% 		$J\mathcal{DQ}_E \gets l'$ +\!\!\!+  $r'$ +\!\!\!+ $counter$\\
				% 		$JQ^l_{E}.add$($JQ^l_{E_{k}}$)\\
				% 		$counter$++
				% 	}
				% 	$JL_E.add$($J$)\\
				% 	$jId \gets counter$
				% }
 			}
		}
		\Return{($J\mathcal{DR}_E$)}
	}
\end{algorithm}

\textbf{Cost Analysis.} The \textit{Deduplicate-Join Operator} implements the Deduplicate operator and two joins; one between a dirty right(left) and a clean left(right) entity sets; $\mathcal{DR}_E$$\bowtie$$QE_E$ followed by a join between the two clean sets $L\mathcal{DR}_E$$\bowtie$$R\mathcal{DR}_E$. The cost of the Deduplicate operator was presented in Sec. \ref{Deduplicate Operator}. The costs of these hash joins are  $3$($|L\mathcal{DR}_{E}|$+$|R\mathcal{DR}_{E}|$) and $3$($|\mathcal{DR}_{E}|$+$|QE_{E}|$), respectively.
\subsection{Group-Entities Operator} \label{Group-Entities Operator}
The \textit{Group-Entities Operator} aims at grouping the results of the above two operations into a single record per entity, before the final \textit{Project}. The operator takes as input a $\mathcal{DR}_E$ and provides as output a grouped set $\mathcal{DR}_G$ containing a single record for each set of duplicate entities. It acts as an aggregate function that groups all attribute values $\forall e_i \equiv e_j$, by concatenation. In this work, we do not focus on any of the data merging techniques for fusing the matching entities, but instead we group them to simplify the presentation of the final projection. For example, if an attribute is written as ``EDBT'' and ``International Conference on Extending Database Technology'' on two matching entities, we create a "hyper-entity" that has [EDBT $|$ International Conference on Extending Database Technology] as value for this attribute (nulls are mapped to an empty value).\silence{This operator acts like a grouping function\footnote{E.g., Impala Query engine (\url{https://bit.ly/3nT8YPJ})  proposes the GROUP-CONCAT function that acts on separate attributes, whereas our operator acts on every attribute of a recordset.} and is executed after all operators have finished and produces a $\mathcal{DR}_G$.}
\silence{This grouping mechanism is presented in Algorithm \ref{alg:group} and is described below.
\begin{algorithm}
\caption{Group-Entities Operator}
\label{alg:group}
\scriptsize
\SetKwInOut{Input}{Input}
\SetKwInOut{Data}{Data}
\Input{
$DR_{E}$
}
\Data{$checkedEntities$ // Set of entities that have been checked
\Output{$DR_G$} // Grouped entities}
\For{$Q_{E_{k}} \in DR_{E}$}{
    % \If{$Q_{E_{k}} \notin checkedEntities$}{
    %     $checkedEntities.add$($Q_{E_{k}}$)\\
        \tcc{Initialize new hyper-entity}
        $DQ_{G_{k}} \gets   Q_{E_{k}}$\\
        \For{$\overline{Q_{E_{k}}} \in DQ_{E}.L_{E_{k}}$}{
        % $checkedEntities.add$($\overline{Q_{E_{k}}}$)\\
            \For{$attribute \in \overline{Q_{E_{k}}}$}{
                \If{$DQ_{G_{k}}$[$attribute$] not empty}{
                     $DQ_{G_{k}}$[$attribute$] $\gets $ " | " + $attribute$
                 }
                 \Else{              $DQ_{G_{k}}$[$attribute$]$     \gets attribute$
                }
               
            }
        }
        $DQ_G \gets DQ_G_k$\\
    % }
}

\Return {$DQ_G$}
\end{algorithm}
The Algorithm \ref{alg:group} iterates over all resolved entities. For each entity it creates a GroupedEntity based on it and its similarities (lines 4,5). If two or more similar entities have a different way of expressing the same attribute, then in the GroupedEntity that is created we concatenate these Strings using a pipe "|" character between them.(lines 9-11). If the values are the same we only add the attribute one time. Each GroupedEntity is added to a final IEnumerable object called GroupedEnumerable}
\section{Dedupe Query Evaluation} \label{ER Query Evaluation}
In this section, we present the methods for planning and evaluating a \textit{Dedupe} query.\silence{in order to correctly answer a query over dirty data while performing the minimal amount of cleaning steps.}
We show how the operators are used to form a query plan by providing two solutions: a naive one which considers fixed plans and an advanced one that selects the plan that minimizes the cost of the comparison execution. We consider \textit{flat SQL} queries which can be represented by a SPJ query tree.\silence{The leaves of a SPJ tree are the relations and the non-leaf nodes are the relational operators (e.g. selection, projections, joins). Each non-leaf node encapsulates one or multiple operations that required to evaluate the query trees. The leaves on the other hand represent the data flow from bottom to top.}
\silence{\subsection{Query evaluation overview}
The evaluation of a query is depicted in Fig. \ref{fig:evaluation}. A user query, is first parsed by the query parser. Then, the planner will create multiple query tree plans and choose the least expensive one. Then, it will pass this plan to the executor which will load the data from the data file and execute the plan. Finally, it will return to the user the results of the query. The evaluation of a query in our approach comprises the following high level steps: 

 \textit{Blocking and Indexing}. This first step creates two indexes per entity collection $E$, which are used in query processing. The $TBI_E$ index, which maps blocks to entities and the inverse index, $ITBI_E$, which maps entities to blocks. $TBI_E$ is sorted in ascending order by the size of its blocks. The blocks of each entity in the $ITBI_E$, are also sorted in ascending order by their respective size. \silence{utilize it also at the statistics phase to estimate the number of comparisons}
 
 \textit{Query Planning}. When a user issues an SQL query, the query engine creates an initial plan without considering the ER process. Consequently, the initial plan is transformed with \textit{operators} insertions and substitutions to what is considered a Dedupe query plan, in a two step approach. In the first step we collect precomputed statistics (e.g. selectivity, estimated number of comparisons, estimated number of joins, etc.). In the second step the planner utilizes these statistics to devise the best plan. 
 
\textit{Query Plan Execution}. Finally, the query engine executes the plan utilizing the indices and the \textit{operators} to produce the final output $\mathcal{DR}_G$.
}
\subsection{Na\"{i}ve ER Solution} \label{Naive ER Solution}
The na\"{i}ve way to answer the SPJ query of the motivating example is to first perform blocking on the entire tables \textit{P} and \textit{V}, process the blocks, perform the comparisons, and finally create a new set of deduplicated entities for answering the query. This is the equivalent of placing the \textit{Deduplicate Operator} directly above the \textit{Table Scan} operators (Fig.\ref{diag:erplan}). However, this solution is expensive (similar to the batch approach), as it will have to clean the whole table prior to evaluating the filter of the query\silence{which is.  on top of the cleaned data while the user is interested only in a subset of both tables} 
(e.g. p.venue="EDBT").\silence{The filter however restricts the selection set, thus publications that do not satisfy that criterion could be eliminated from the deduplication process.} 
An obvious plan enhancement would be to put the \textit{Deduplicate Operator} above the filter on the left branch of the tree (Fig. \ref{diag:erplan2})\silence{Placing the \textit{Deduplicate Operator} above the \textit{Filter Operator} makes it feasible to} 
hence reducing the number of entities $|QE_E|$ that will initially feed the \textit{Deduplicate Operator}. %that are going to be later cleaned, by taking as input a subset of the table, $Q_E \subseteq E$. 
Finally, we observe that $V$ will not join with all of the cleaned publications. Thus, eliminating those venues that do not join with the  publications prior to cleaning them, could further reduce the computational cost. 
\silence{Based on these intuitions we propose an advanced solution for integrating ER with the query processing in the next section.}
\begin{figure}[h]
\centering
\begin{minipage}[t]{.2\textwidth}
    \tiny
    \Tree[.{\textit{Project}}
    [.{\textit{Join}}  
    [.{ \textit{$Filter_{venue='EDBT'}$}}
    [.{ \textit{Deduplicate}}  
    [.{ \textit{Table scan}} 
    [.{\textit{P}}
    ]]]]
    [.{ \textit{Deduplicate}}  
    [.{\textit{Table scan}}
    [.{\textit{V}}
    ]
    ]]]]
    \caption{Naive ER plan}
    \label{diag:erplan}
\end{minipage}
\hspace{1.2mm}
\begin{minipage}[t]{.2\textwidth}
    \tiny
    \Tree[.{\textit{Project}}
    [.{\textit{Join}}  
    [.{ \textit{Deduplicate}}
    [.{ \textit\tiny{$Filter_{venue='EDBT'}$}}
    [.{ \textit{Table scan}} 
    [.{\textit{P}}
    ]]]]
    [.{ \textit{Deduplicate}}  
    [.{\textit{Table scan}} 
    [.{\textit{V}} ]]]]]
    \caption{Naive ER plan 2}
    \label{diag:erplan2}
\end{minipage}
\end{figure}

\subsection{Advanced ER Solution} \label{Advanced ER Solution}
Based on the above observations, we present a \textit{cost-based} approach that attempts to significantly improve the performance of the query, by reducing the unnecessary comparisons in each table before the computationally expensive \textit{Comparison-Execution}.\silence{The key concepts of this approach are the \textit{operators} defined in section \ref{Operators} as well as the \textit{Statistics} defined in this section.}
\subsubsection{Query Planning} \label{Query Planning}. 
The \textit{Advanced ER Solution} assumes that the best non ER-enabled query plan that contains the best 
operators placement is given. For instance, the initial input to the \textit{Advanced ER Solution} is a plan 
similar to the plan in Fig. \ref{diag:qplan}. We convert this plan into a plan of Fig. \ref{diag:splan1} or 
\ref{diag:splan2} (depending on the statistics) by inserting the \textit{Deduplicate Operator} and the 
\textit{Group-Entities Operator} and substituting the initial Join Operator with the \textit{Deduplicate-Join 
Operator}. The operator placement is done with the purpose of eliminating as much unnecessary comparisons as 
possible before the \textit{Comparison-Execution} step while ensuring the correctness of the final output as 
well as to devise an optimal \textit{Join Ordering} when multiple joins are involved. Query Planning works as 
follows:

i) \textit{Statistics}. The first step of the approach is to compute the query statistics; i.e., the estimated number of comparisons and the estimated number of join predicates. 

To estimate the number of comparisons of a query, we utilise the \textit{WHERE} clause. For the estimation, we consider that a literal used in a condition expression defines a Blocking Key ($BK$) in the Table Block index 
$TBI_E$. Let $WB$ be the subset of blocks that their $BKs$ are contained in the \textit{WHERE} clause as 
literals. Then, for each block $b_{i}\in WB$, we get the corresponding block $b_i \in TBI_E$. Based on 
the operators $AND$, $OR$ of the clause, we derive the estimated selected set $S_E \approx QE_E$ that consists of all the disjuncted and/or conjucted entities $e$ that belong in blocks of $WB$. Now, $ \forall e \in 
S_E \setminus LI_E$, we get all its corresponding blocks from $ITBI_E$ and create the block collection $SB$ 
to approximate the $EQBI_E$.
Next, we apply the block purging and filtering algorithms to approximate the number of comparisons after the meta-blocking step. For purging, we heuristically identify oversized blocks by estimating an upper limit on the number of matches a block is expected to contain. For that, we need to calculate the comparison threshold $t \leq$ MAX($||b_i||$) where a block $b_i$ is removed if $||b_i|| > t$. This number will be equal to $||b_i||$ where $|b_i| \cdot ||b_{i - 1}|| < SF \cdot ||b_{i}|| \cdot |b_{i - 1}|$, where  \textit{SF} is a smoothing factor experimentally set to 1.025 \cite{Papadakis13}. \silence{ by targeting an increase in efficiency with limited impact on effectiveness.}
To approximate the number of filtered comparisons, we utilize the $ITBI_E$ where the blocks (values of $ITBI_E$) are pre-sorted in ascending order by their size $|b|$. So, $\forall$ $e_i$$\in$$ITBI_E$ we take the set of blocks $\{B\}$ that it belongs to and retain the $e_i$ only in the first $n$ blocks where $n=p$$\cdot$$|\{B\}|$ and $p$$\leq$$1$ is the filtering parameter. Because the blocks are pre-sorted in ascending order by their size $|b|$, we can easily remove the entities from the blocks that are above the threshold.
The final number of estimated comparisons for this table is, $C = \Sigma_{S_{b_i} \in SB}|q_b| \cdot$ 
($|S_{b_i}|$ - $(|q_b|+1)/2$), where $q_b$ is the set of entities $\in$ $S_E \setminus LI_E$ within each 
$S_{b_i} \in SB$. The purpose of the comparisons estimation is to estimate which of both tables, that take 
part in the join, is yielding the highest number of comparisons and delay its deduplication. Since the cost of estimating the output of the \textit{Edge Pruning} (discussed in Sec.\ref{Deduplicate Operator}) is very high, we terminate our calculations at the $BF$ step, where a safe conclusion about the inequality can be drawn.

To estimate the number of Join predicates, we need to first estimate (i) the size of $\mathcal{DR}_E$ and (ii) the size of the $\mathcal{DR}_E$ for each table after the join. For the estimated $|\mathcal{DR}_E|$, a sample of each table is eagerly cleaned offline, during the initial data loading. From that, we calculate the duplication factor, $df$, i.e., the approximate number of duplicates it contains. For example, if the sample $S$ of a table with $|S| = |QE_E| = 800$ entities, has  $|\mathcal{DR}_E| = 1000$ entities then the estimated percentage of duplicates in that tables is $20\%$. Thus, for a query on the aforementioned tables that yields a $|QE_E|=2000$ entities the estimated size of $\mathcal{DR}_E$   will be 2400 entities. 
To estimate the size of the $\mathcal{DR}_E$ after the join, we pre-compute for every table pair the percentage of entities that join. With this, we can calculate the reduction of each $\mathcal{DR}_E$ after the join. For instance, if we know that $20\%$ of $T1$ joins with $50\%$ of $T2$ we can estimate that their respective $\mathcal{DR}_E$ sizes will be reduced by the same amount. \silence{Finally, the estimated cardinality of the Join result is calculated from the product of the estimated sizes  $\mathcal{DR}_E$ of the two tables $JP =  $}

ii) \textit{ER Query Planning}. The planner in this solution decides the best placement of the operators that will ensure the $\mathcal{DQ}$ Performance.\silence{Since the \textit{Comparison-Execution} dominates the execution time the later goal can be translated to the minimization of the \textit{total number} of pair-wise comparisons that a query will perform. } The query plan for the \textit{SP} class of queries is straightforward since the planner cannot change many things. In such cases, the \textit{Deduplicate Operator} is added on top of the \textit{Filter}. Placing the operator above the \textit{Filter} reduces the number of entities $|QE_E|$ that will initially feed the \textit{Deduplicate Operator}. In cases the \texttt{WHERE} clause is not present, we have to deduplicate the whole table.
For the \textit{SPJ} class of queries one branch of the query tree has to be deduplicated prior to the Join to ensure the \textit{Correctness}. The planner utilizes the statistics and places the \textit{Deduplicate Operator} to the branch that yields the lowest number of comparisons. Note that  %For the reason that the \textit{Deduplicate-Join Operator} first forms a \textit{QBI} for the "dirty" entity-set, from the entities that join with the cleaned one (\textit{filter-join}), 
the total number of comparisons that will occur after the \textit{Deduplicate-Join} will be lower if we first deduplicate the branch that yields the lowest number of comparisons\silence{ than the other way around (the entity-set from the other branch will be further restricted)}. This is due to the reason that the \textit{Deduplicate-Join Operator} first forms a \textit{QBI} for the "dirty" entity-set, from the entities that join with the deduplicated one. Based on this, the planner decides the appropriate type of \textit{Deduplicate-Join Operator}. For instance, the plan in Fig.\ref{diag:splan2} will be chosen as best because the total number of comparisons (Table \ref{tbl:clean-first}) is less than the one in the plan of Fig.\ref{diag:splan1}.\silence{ This is because, cleaning table $V$ initially yields 15 comparisons and cleaning table $P$ (after the filter) yields 18.}
%\RED{krypsto , den hmaste sigouroi k topetame etsi}The aforementioned case stands only for the first \textit{Join}, when the number of joins > 1, since after the first join the sum of all comparisons that will be executed during the query process will be the same independently from the \textit{Join} order. 
The planner optimises the join order\silence{ (after the first join)} based on the statistics, to choose the plan that ensures the minimum I/O and memory utilization. Finally, the \textit{Group-Entities Operator} is placed directly before the final \textit{Project} to form the deduplicated grouped entities. %for the simplification and the better visualisation of the final result-set.
 \begin{table}[t]
\centering
\tiny
\setlength\extrarowheight{0.99mm}
% \small
\begin{tabular}{||l|l|l|l||}
\hline
\textbf{Clean First} &  \multicolumn{2}{l|}{\textbf{Comparisons}} & \textbf{Total}\\ 
&  $V$ & $P$ & \\
\hline
$V$ & 12 & 3 & 15\\
$P$ & 17 & 1 & 18\\
\hline
\end{tabular}
\caption{Exec. Comp. based on Cleaning Order}
\label{tbl:clean-first}
 \end{table}
\subsubsection{Query Execution} \label{Query Plan Execution}. 
The solution executes the plan based on the ordering that the planner assigns to each operator. \framework utilizes the established database pipelining architecture where the output of an operator is passed to its parent \silence{without the materialization of the intermediate results }by implementing the Iterator Interface. For example, given that the input to the executor will be the plan of  Fig. \ref{diag:splan2}, %. The plan will be executed in the following order:
the evaluation will start from the right branch of the query tree since it first has to evaluate the \textit{Deduplicate Operator}. First, the table $V$ will be scanned and subsequently the \textit{Deduplicate Operator} is evaluated. 
\silence{The Operator first creates the \textit{QBI} using the entities $Q_E$ from the \textit{Table Scan} and then performs the \textit{Block-Join Operation} between the \textit{QBI} and the \textit{TBI} to enhance the formerly created blocks with all the possible duplicates of the retrieved entities. After the \textit{Block-Join Operation}, the \textit{Meta-Blocking Operation} is performed on the $QBI$ to reduce its size $|QBI|$ and cardinality ||$QBI||$. Finally, the output of the \textit{Meta-Blocking Operation} is used by the \textit{Comparison-Execution Operation} to resolve the duplicates and produce the $\mathcal{DQ}_E$. } Then, it executes the \textit{Table Scan} and the \textit{Filter} on table $P$ to retrieve all entities that satisfy the user's query. Next, the \textit{Deduplicate-Join Operator} is evaluated.\silence{In our example, we have a \textit{Dirty-Left} join since the left branch will be deduplicated by the \textit{Deduplicate-join operator}.}

\silence{Specifically, the operator first will create a \textit{QBI} from the output of \textit{filter-join} for the left dirty entity-set, to further reduce its $|Q_E|$. The \textit{filter-join} of the two entity-sets, in such cases, acts like the \textit{Filter Operation} and allows us to discard the entities that do not join and thus to clean only a subset of the whole dataset. Finally it will perform a \textit{Deduplicate operation} on the $QBI$ of the dirty side.}

\begin{figure}[h]
\begin{minipage}[t]{.2\textwidth}
\tiny
\centering
\Tree[.\textit{Project}
     [.{\textit{GroupEntities}}
    [.{\textit{Dirty-RightJoin}}
    [.{ \textit{Deduplicate}}
    [.{ \textit{$Filter_{venue='EDBT'}$}} 
    [.{\textit{Table scan}} 
    [.{\textit{P}} ]]]]
    [.{ \textit{Table scan} } 
    [.{\textit{V}} ]]]]]
    \caption{Adv. Solution Plan 1}
    \label{diag:splan1}
\centering
\end{minipage}
\hspace{1.2mm}
\begin{minipage}[t]{.2\textwidth}
\tiny
\centering
\Tree[.\textit{Project}
    [.{\textit{GroupEntities}}
    [.{\textit{Dirty-LeftJoin}}   
    [.{ \textit{$Filter_{venue='EDBT'}$}} 
    [.{\textit{Table scan}} 
    [.{\textit{P}} ]]]
    [.{ \textit{Deduplicate}}
    [.{ \textit{Table scan} } 
     [.{\textit{V}} ]]]]]]
    \caption{Adv. Solution Plan 2}
    \label{diag:splan2}
\centering
\end{minipage}
\end{figure}
% \vspace{-10pt}
Next, the \textit{Group-Entities Operator is evaluated}. The operator takes as input the output of the previous operator \silence{\textit{Deduplicate-Join Operator} (or the \textit{Deduplicate Operator if no Join exists)} }and iterates over all deduplicated entities. For each entity it creates a new \textit{GroupedEntity} including the attribute values of all its duplicates. At the last step we have the final \textit{Projection} of the result-set. 
\section{Related Work} \label{Related Work}
\textbf{Entity Resolution.} Given its importance, Entity Resolution has been studied thoroughly from the database community \cite{Ipeirotis07,Lenzerini02,Papadakis21}. Due to their quadratic complexity, existing ER approaches typically scale to large datasets through blocking methods which in principal compare only similar entities. Unlike the exhaustive ER techniques, blocking offers an approximate solution, sacrificing some recall in order to enhance precision. Primarily, the existing methods pertain to structured data, which abide by a specific schema with known semantics and qualitative characteristics for each attribute (Schema-based blocking) \cite{Pchris122}. However, this approach is not applicable to Web Data, due to their highly heterogeneity. For that, blocking methods have been extended to function independent from the schema (schema-agnostic blocking), where every token from every value of every entity is treated as blocking key \cite{Alexiou15}. Although this approach successfully tackles the heterogeneity, it creates overlapping blocks resulting in unnecessary comparisons \cite{Alexiou15,Papadakis133}. Therefore, they must be avoided. This is achieved by block processing techniques that are appropriate for Web Data as they successfully tackle the heterogeneity and the great volume of the data. The most important method is the Meta-blocking \cite{Papadakis133, Papadakis16} which eliminates all the redundant and the superfluous comparisons by examining the block-to-entity relationships. A notable work that tries to tackle the problem of ER in heterogeneous Web Data is MinoanER \cite{Efthymiou16}, which relies in Schema-agnostic techniques that consider the content and the neighbors of the entities, but it does it in a batch/offline processing. Instead, our approach operates online during query time.

\textbf{Analysis-aware cleaning.} Over the last years, a few methods for integrating Entity Resolution with Query Processing have been proposed with the aim to answer SQL-like queries  over erroneous data. Some of these methods are approximate solutions that are not designed for the larger class of SPJ queries \cite{Altwaijry13}, or do not even consider optimizing for other types of selection queries such as range queries or queries where the type of the condition attribute is not a string \cite{Bhattacharya07}. The approach in \cite{Alexiou19} only considers simple SP queries on single entity collections while it operates on top of existing query engines thus disregarding query planning and optimisation procedures. Other approaches are only considering probabilistic databases, which assume the existence of link-sets between entities. More specifically, the approaches in \cite{Ioannou10, Andritsos06, Sismanis09} are focusing on answering simple single-table aggregation queries, while in \cite{Ioannou15}, on a range of topK and Iceberg aggregation queries by joining multiple tables. Also, Sample-and-clean \cite{Wang14} extracts a sample from a dataset with duplicates and uses the sample to answer aggregate queries while considering data cleaning as a user-provided module. Sample-and-clean estimates the query's answer given the user's input and corrects the error of the queries over the uncleaned data. \silence{Sample-and-clean requires the user intervention in order to clean the data whereas our approach does not require none of the above.}
Nevertheless, a handful of notable exceptions exist. 

CleanDB \cite{Giannakopoulou@vldb17}, integrates deduplication with query processing, in a distributed setting, by cleaning the whole table instead of cleaning only the parts of the data that are needed by a given query. Therefore, CleanDB addresses a different dimension of the scalability issue of data cleaning than \framework does.
Daisy\cite{Giannakopoulou@sigmod20} is a system that performs probabilistic repair of functional dependency violations with query-result relaxation that enables interleaving SPJ queries. Daisy also introduces update operators which differentiate between Select and Join operators, inside the query plan by analyzing the query operators affected by the constraints. In cases of SPJ queries it first cleans both tables and then re-executes the join operator. Daisy focuses on integrity constraints while we are focusing on ER techniques. The most relevant work to our approach is QuERy \cite{Altwaijry15}. QuERy is a framework which enables the evaluation of SPJ queries over data with duplicates. It uses blocking for processing and requires that data are first pre-processed and grouped together in corresponding blocks along with their summaries (sketches). It is offered in two variants: i) a lazy-solution that does not consider costs and, ii) an adaptive-solution that uses a cost-based planner. Similarly to our approach, it introduces operators in the query plan that operate over these blocks; however, neither the source code nor their datasets are publicly available for performing a direct experimental comparison. Nevertheless, as a qualitative comparison in performance issues, we can observe from the experimental results reported in \cite{Altwaijry15} (Fig.19) that the execution time for all their provided solutions (SPJ queries with low selectivity on $C_{|80070|}\bowtie M_{|1237|}$) is over 100 sec. while in \framework for queries with similar selectivity the time is below 100 sec. even for $OAGP_{2M}$$\bowtie$$OAGV_{130K}$ (Fig.\ref{nes_vs_aes_time_scale_2}). This significant difference could be explained by the fact that \framework does not need to perform neither sampling and calculation of statistics nor eager cleaning of blocks for every issued query to determine the best possible solution. Instead, it uses the block-join operation to retrieve all the possible "dirty" subsets that answer the user's query and subsequently it utilizes Meta-blocking to reduce the cleaning overhead instead of block sketches that seem to work well only on datasets with many numerical values.

\textbf{Similarity Joins.} A related approach to ER is the Set Similarity Joins (SSJ) which compute all pairs of similar sets from two collections of sets. A recent survey \cite{Fier18} included top works in this area and reported on ten recent, distributed set similarity join algorithms with some interesting results: All algorithms in their tests failed to scale for at least one dataset and were sensitive to frequent set elements, low similarity thresholds, or a combination thereof. Moreover, some algorithms even failed to handle the small datasets that can easily be processed in a non-distributed setting. SSJ can be considered an orthogonal setting to ER, since it focuses on improving the performance of distance functions in join operations, rather than resolving duplicate records.

\pgfplotsset{
    /pgfplots/ybar legend/.style={
    /pgfplots/legend image code/.code={%
       \draw[##1,/tikz/.cd,yshift=-0.25em]
        (0cm,0cm) rectangle (3pt,0.8em);},
   },
}
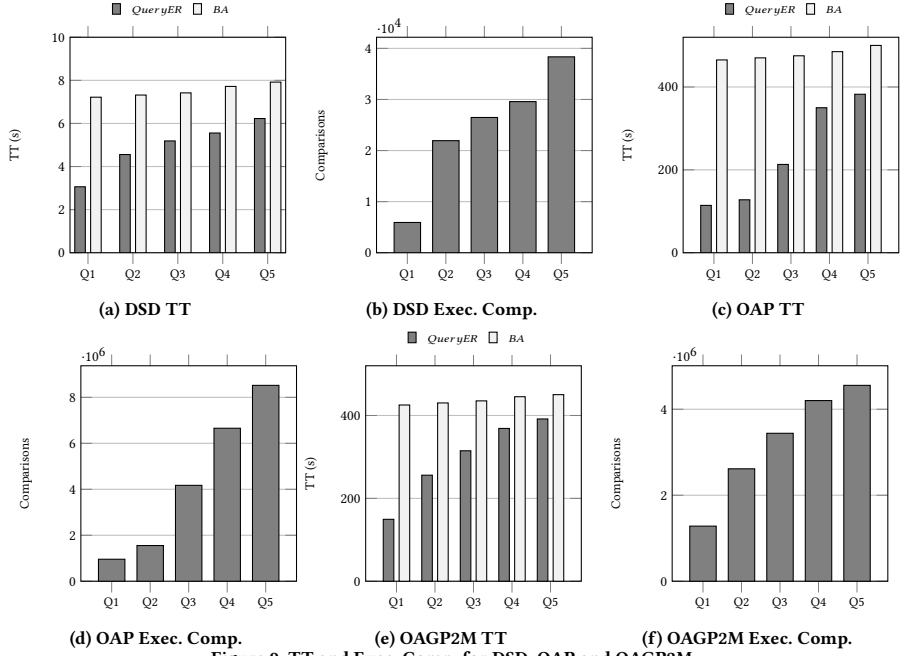
\begin{figure*}[h!]
\tiny
\centering  
\subfloat[DSD TT]{%
\centering
\begin{tikzpicture}
\begin{axis}[width=0.25\linewidth,
 height=0.25\linewidth,
ylabel=TT (s), 
ylabel style={font=\tiny,  yshift=-5mm},
xlabel style = {font=\tiny},
yticklabel style = {font=\tiny},
xticklabel style = {font=\tiny},
bar width=4,
legend entries={$QueryER$, $BA$},
legend columns=-1,
enlarge x limits=0.1,
legend style={font=\fontsize{4}{5}\selectfont,
                at={(0.8,1.05)},
                anchor=south east,
                column sep=1ex,
                draw=none
        },
ymin=0,ymax=10,
ybar ,
xtick=data,
symbolic x coords={Q1, Q2, Q3, Q4, Q5},
grid=major,
xmajorgrids=false]
\addplot 
[fill=gray!100]
	coordinates {
	({Q1}, 3.06)
	({Q2}, 4.5576)
	(Q3, 5.1884)
	(Q4, 5.556)
	(Q5, 6.2276)};
\addplot 
 [fill=gray!10]
	coordinates {
    ({Q1}, 7.218)
	({Q2}, 7.318)
	(Q3, 7.418)
	(Q4, 7.718)
	(Q5, 7.918)};
% \legend{TT}
\end{axis}
\end{tikzpicture}
\label{dsd-toa}}
\hspace{1.4em}
\subfloat[DSD Exec. Comp.]{%
\begin{tikzpicture}
\begin{axis}[
 width=0.25\linewidth,
       height=0.25\linewidth,
ylabel=Comparisons,
ylabel style={font=\tiny,  yshift=-5mm},
xlabel style = {font=\tiny},
yticklabel style = {font=\tiny},
xticklabel style = {font=\tiny},
enlarge x limits=0.2,
legend style={font=\fontsize{4}{5}\selectfont,
                at={(1,1.05)},
                anchor=south east,
                column sep=1ex
        },
ymin=0,
ybar,
xtick=data,
symbolic x coords={Q1, Q2, Q3, Q4, Q5},
grid=major,
xmajorgrids=false
]

\addplot 
[fill=gray!100]
	coordinates {
	({Q1}, 5922)
	({Q2}, 21936)
	(Q3, 26475)
	(Q4, 29582)
	(Q5, 38343)};
\end{axis}
\end{tikzpicture}
\label{DSD-comps}}
\hspace{1.4em}
\subfloat[OAP TT]{%
\begin{tikzpicture}
\begin{axis}[
 width=0.25\linewidth,
 height=0.25\linewidth,
ylabel=TT (s),
ylabel style={font=\tiny,  yshift=-5mm},
yticklabel style = {font=\tiny},
xticklabel style = {font=\tiny},
bar width=4,
enlarge x limits=0.2,
legend style={font=\fontsize{4}{5}\selectfont,
        at={(0.8,1.05)},
        anchor=south east,
        draw=none,
        column sep=1ex,
        },
legend entries={$QueryER$, $BA$},
legend columns=-1,
ymin=0,ymax=520,
ybar ,
xtick=data,
symbolic x coords={Q1, Q2, Q3, Q4, Q5},
grid=major,
xmajorgrids=false
]

\addplot
[fill=gray!100]
	coordinates {
	({Q1}, 114.26)
	({Q2}, 127.7678)
	(Q3, 213.112)
	(Q4, 349.9188)
	(Q5, 382.4638)};
\addplot
[fill=gray!10]
	coordinates {
	({Q1}, 465.2872)
	({Q2}, 470.2872)
	(Q3, 475.2872)
	(Q4, 485.2872)
	(Q5, 500.2872)};
\end{axis}
\end{tikzpicture}
\label{oap-toa}}
\hspace{1.4em}
\subfloat[OAP Exec. Comp.]{%
\begin{tikzpicture}
\begin{axis}[
 width=0.25\linewidth,
       height=0.25\linewidth,
ylabel=Comparisons,
ylabel style={font=\tiny,yshift=-5mm},
xlabel style = {font=\tiny},
yticklabel style = {font=\tiny},
xticklabel style = {font=\tiny},
enlarge x limits=0.2,
legend style={ font=\fontsize{4}{5}\selectfont,
                at={(1,1.05)},
                anchor=south east,
                column sep=1ex
        },
ymin=0,
ybar,
xtick=data,
symbolic x coords={Q1, Q2, Q3, Q4, Q5},
grid=major,
xmajorgrids=false
]

\addplot [fill=gray!100]
	coordinates {
	({Q1}, 957381)
	({Q2}, 1553186)
	(Q3, 4171684)
	(Q4, 6655850)
	(Q5, 8521269)};
\end{axis}
\end{tikzpicture}
\label{oap-comps}}
% \label{xp:SP1}
\subfloat[OAGP2M TT]{%
\begin{tikzpicture}
\begin{axis}[
 width=0.25\linewidth,
 height=0.25\linewidth,
ylabel=TT (s),
ylabel style={font=\tiny,  yshift=-5mm},
yticklabel style = {font=\tiny},
xticklabel style = {font=\tiny},
bar width=4,
enlarge x limits=0.2,
legend style={font=\fontsize{4}{5}\selectfont,
        at={(0.8,1.05)},
        anchor=south east,
        draw=none,
        column sep=1ex,
        },
legend entries={$QueryER$, $BA$},
legend columns=-1,
ymin=0,ymax=520,
ybar ,
xtick=data,
symbolic x coords={Q1, Q2, Q3, Q4, Q5},
grid=major,
xmajorgrids=false
]

\addplot
[fill=gray!100]
	coordinates {
	({Q1}, 149.447)
	({Q2}, 255.936)
	(Q3, 314.848)
	(Q4, 368.763)
	(Q5, 391.651)};
\addplot
[fill=gray!10]
	coordinates {
	({Q1}, 425.2872)
	({Q2}, 430.2872)
	(Q3, 435.2872)
	(Q4, 445.2872)
	(Q5, 450.2872)};
\end{axis}
\end{tikzpicture}
\label{oagp2m-toa}}
\hspace{1.4em}
\subfloat[OAGP2M Exec. Comp.]{%
\begin{tikzpicture}
\begin{axis}[
 width=0.25\linewidth,
       height=0.25\linewidth,
ylabel=Comparisons,
ylabel style={font=\tiny,yshift=-5mm},
xlabel style = {font=\tiny},
yticklabel style = {font=\tiny},
xticklabel style = {font=\tiny},
enlarge x limits=0.2,
legend style={ font=\fontsize{4}{5}\selectfont,
                at={(1,1.05)},
                anchor=south east,
                column sep=1ex
        },
ymin=0,
ybar,
xtick=data,
symbolic x coords={Q1, Q2, Q3, Q4, Q5},
grid=major,
xmajorgrids=false
]

\addplot [fill=gray!100]
	coordinates {
	({Q1}, 1281911)
	({Q2}, 2614897)
	(Q3, 3441096)
	(Q4, 4201777)
	(Q5, 4555969)};
\end{axis}
\end{tikzpicture}
\label{oagp2m-comps}}
\caption{TT and Exec. Comp. for DSD, OAP and OAGP2M}
\label{xp:SP1}
\end{figure*}
\begin{table}[h]
\centering
\footnotesize
\begin{tabular}{||l|l|l|l|l|l|l|l|l|l|l|||}
\hline
%\textbf{E} & \textbf{TT}  &  \textbf{BJT} & \textbf{MBT} & \textbf{RT}  & \textbf{GET} & \textbf{Other} \\ 
\textbf{E} & \textbf{TT (s)}  &  \textbf{\vtop{\hbox{\strut Block-}\hbox{\strut Join}}} & \textbf{\vtop{\hbox{\strut Meta-}\hbox{\strut blocking}}} & \textbf{Resolution}  & \textbf{Group} & \textbf{Other} \\ 
\hline
DSD & 6.2274 & 7\% &  5\% & 82\% & 3\%  & 3\% \\
OAP & 422.46 & 5\% &  7\% & 83\% & 1\% & 4\% \\
\hline
\end{tabular}
\caption{ TT breakdown on DSD and OAP for Q5.} 
\label{tbl:time_breakdown}
\end{table}
\vspace{-20pt}
\section{Experimental Evaluation} \label{Experimental Evaluation}
We experimentally evaluated the effectiveness, the efficiency and the scalability of our approach on several real and synthetic datasets.

\subsection{Experimental Setup}
We have implemented \framework in Java version 8. The experiments were performed on a desktop computer with Intel i7 (3.4GHz) and 64GB of RAM. All measurements were repeated 10 times and the average value is reported. All the resources as well as a link for an online demo, are available in GitHub\footnote{\url{https://github.com/VisualFacts/queryER}}.

\textbf{Datasets}. Our experimental analysis involves (i) real, (ii) real that have been modified to include duplicates and (iii) synthetic datasets. It is worth mentioning that, while there exist plenty of datasets for evaluating ER workflows, almost none of them can be used to perform SPJ queries. To this end, we had to manually modify real datasets to include duplicates.\par

$\bullet$ The \textit{DBLP-Scholar}\cite{Kopcke10} (\textbf{DSD}) is an established dataset that has been widely used in literature \cite{Kopcke10, Efthymiou16}. It contains bibliographic records from DBLP and Google Scholar. The Open Academic Graph \cite{Sinha15, Tang08} (\textbf{OAG}), which is a large knowledge graph unifying two billion-scale academic graphs: MAG\footnote{\url{https://academic.microsoft.com}} and AMiner\footnote{\url{https://www.aminer.cn}}. Specifically, we used the Paper data (\textbf{OAGP}) and Venue data (\textbf{OAGV}). For OAGP, we created 5 different size variations (\textbf{OAGP200K-2M}), for scalability testing (ground-truth provided). 

$\bullet$ The \textit{Organisations} (\textbf{OAO}) and \textit{Projects} (\textbf{OAP}) datasets are real datasets that have been obtained from OpenAIRE project \cite{openaire}. They contain records for research organizations and the corresponding projects that these organisations participate in. Both datasets have been modified using the \textit{febrl} \cite{Pchris02} to include $10\%$ duplicate records.

$\bullet$  The synthetic datasets \textit{People200K-2M} (\textbf{PPL200K-2M}) were also created using the \textit{febrl}. First, duplicate-free, people records were produced based on frequency tables of real-world data. Also, an extra attribute was explicitly added to each record to assign an organisation to each person (from OAO) to create a relashionship between them. Then, duplicates of these records were randomly generated based on real-world error characteristics. The resulting datasets contain $40\%$ duplicate records with up to 3 duplicates per record, no more than 2 modifications/attribute, and up to 4 modifications/record. Table \ref{tbl:datasets} summarizes their technical characteristics.
 
\textbf{Evaluation Measures.} The effectiveness and the efficiency of our approach, are assessed using the following measures: 
(\textbf{a}) \textit{Pair Completeness (PC)} estimates the effectiveness (recall) of $EQBI_{QE_E}$ after the \textit{Meta-Blocking Operation} based on a ground-truth $GT$, i.e., the portion of duplicates $\mathcal{D}$ from the input $QE_E$ that co-occur in at least one block of $EQBI_{QE_E}$. More formally, PC = $\mathcal{D}(EQBI_{QE_E})$ / $GT(EQBI_{QE_E})$  and is defined in [0, 1]\silence{, with higher values showing higher effectiveness}. 
(\textbf{b}) Total execution time \textit{(TT)} of the query. 
(\textbf{c}) Comparisons \textit{(Comp.)}, which is the number of executed comparisons that yielded from each query and it is a supplementary measure to TT.

\textbf{Baseline and Configurations.} We experimented with the different alternatives of our approach (w/o the planner, w/o the link index $LI$) considering the batch approach $BA$ (defined in Sec.\ref{Problem Statement}) as our baseline. For all configurations we used a fixed blocking (Token-Blocking) and meta-blocking strategy and the \textit{Jaro-Winker} similarity function. To the best of our knowledge, apart from \cite{Altwaijry15}, which was extensively discussed in Sec.\ref{Related Work}, there are no other direct competitors that have implemented similar operators.

\textbf{Evaluation Workload.} To evaluate the performance, the scalability and our cost-based planner, we designed a workload of 13 types of SP and SPJ queries with ranging and fixed selectivity $S$. For each experiment we chose the most appropriate combination of both synthetic and real datasets, based on their size $|E|$ and technical characteristics. Specifically, the workload we use evaluates:\par
\textbf{(a)} the performance of \framework against $BA$ (Fig.\ref{xp:SP1}) on $DSD$, $OAP$ and $OAGP2M$ datasets. For this evaluation, we used 5 SP queries ($Q1-Q5$) per dataset, for each experiment, with $S$ ranging from $\approx5\%$ to $\approx80\%$ with an approximate step $15\%$ to examine the performance of \framework on different dataset types with increasing sizes.

\textbf{(b)} the scalability of \framework over an increasing dataset size with fixed selectivity (Fig.\ref{xp:sp-scal}) using both real ($OAGP200K-2M$) and synthetic ($PPL200K-2M$) datasets. To perform this evaluation, we used $Q9$ = \textit{MOD(id, 10)$<$1} on both datasets to ensure the randomness of the selection along with the fixed $|QE_E|$ with increasing $|E|$.

\textbf{(c)} the effects of the index $LI$ on \framework's performance (Fig.\ref{xp:sp-links}) by executing 4 overlapping range queries ($Q10-Q13$) on the $OAGP2M$ dataset. Each query contains the $QE_E$ of the previous plus 30\% more entities, starting with $Q10$ which has $|QE_E|$ = 760000. %, except $Q10$,

\textbf{(d)} the execution time of each step of the proposed workflow on one real ($DSD$) and on one synthetic ($OAP$) dataset using the SP query with the highest selectivity ($Q5$).

\textbf{(e)} the effects of different \textit{Meta-Blocking} configurations on real ($OAGP1M$) and synthetic ($PPL1M$) datasets (Tbl.\ref{tbl:MB}) using  queries $Q1$ and $Q5$ with the lowest and the highest selectivity respectively.

\textbf{(f)} the performance of our \textit{cost-based planner} by comparing the \textit{Advanced} with the \textit{Na\"{i}ve ER Solution} while using $BA$ as baseline. To this end, we designed a series of SPJ queries on both real and synthetic datasets. In all queries, we performed \textit{joins} between two tables while keeping the selectivity of the one side fixed (100\%). More specifically, we used the following SPJ queries: $Q6_a$ = PPL2M $\bowtie$ OAO, $Q6_b$ = OAGP2M $\bowtie$ OAGV with $S_{PPL2M}$=$S_{OAGP2M}$=$7\%$ and $S_{OAO}$=$S_{OAGV}$=$100\%$, $Q7_a$ = OAP $\bowtie$ OAO, $Q7_b$ = OAGP2M $\bowtie$ OAGV with $S_{OAP}$=$S_{OAGP2M}$=$75\%$ and $S_{OAO}$=$S_{OAGV}$=$100\%$, $Q8_a$ = PPL200K-2M $\bowtie$ OAO, $Q8_b$ = OAGP200K-2M $\bowtie$ OAGV with\\ $S_{PPL200K-2M}$=$S_{OAGP200K-2M}$=$15\%$ and $S_{OAO}$=$S_{OAOGV}$=$100\%$.
\vspace{-4pt}
\begin{table}[h]
\footnotesize
\centering
\begin{tabular}{||l|l|l|l|l|l||}
\hline
\textbf{E} & \textbf{$|E|$}  & \textbf{$|L_E|$} & \textbf{$|$A$|$} & \textbf{$|$TBI$|$} \\ 
\hline
DSD & 66879 & 5347 & 4 & 88K \\
OAO &	55464 & 5464 &	3 & 22K\\
OAP &	500K & 58074 & 8 & 170K\\
PPL200K & 200K & 64762 & 12 & 160K\\
PPL500K & 500K & 161443 & 12 & 280K\\
PPL1M & 1M & 322722 & 12 & 470K\\
PPL1.5M & 1.5M & 403417 & 12 & 590K\\
PPL2M & 2M & 645489 & 12 & 850K\\
OAGP200K & 200K & 5679 & 18 & 110K\\
OAGP500K & 500K & 54132 & 18 & 180K\\
OAGP1M & 1M & 78341 & 18 & 240K\\
OAGP1.5M & 1.5M & 135313 & 18 & 320K\\
OAGP2M & 2M & 267843 & 18 & 360K\\
OAGV & 130K & 29841 & 5 & 55K\\
\hline
\end{tabular}
\caption{$|E|$:dataset size, $|L_E|$:\# of duplicates, $|A|$:\# of distinct attribute names, $|TBI|$:size of the TBI}
\label{tbl:datasets}
 \end{table}\newline
\vspace{-5pt}
\subsection{Performance and Scalability} \label{SP-EXPS}
\textbf{\framework vs \textbf{BA}}. Fig.\ref{dsd-toa},  Fig.\ref{oap-toa} and Fig.\ref{oagp2m-toa} show the $TT$ for \framework vs $BA$ for ranging $S$. We observe that as the selectivity increases, the $TT$ of both approaches increases too. However, \framework exhibits significantly higher performance than the $BA$. The difference is due to the fact that, our \textit{Deduplicate Operator} restricts the number of comparisons required to resolve the duplicate entities by deduplicating only those parts of data that influence the query’s answer. On the other hand, the $BA$ has to clean the whole dataset prior to the query execution, to resolve the duplicate pairs. We also observe that as the selectivity increases, the difference between the two approaches decreases because the parts of data that influence the query’s answer tend to be equal to the size of the whole dataset. From Fig.\ref{DSD-comps}, Fig.\ref{oap-comps} and Fig.\ref{oagp2m-comps} we can also observe that time is strongly correlated and is analogous to the executed comparisons.

\begin{figure}
\subfloat[]{%
\begin{tikzpicture}
\hspace*{-1cm} 
\begin{axis}[
ylabel= TT (s),
legend style={font=\fontsize{4}{5}\selectfont,
        at={(1.1,1.05)},
        anchor=south east,
        draw=none,
        column sep=1ex,
        },
legend entries={$QueryER$ PPL, $BA$ PPL, $QueryER$ OAGP, $BA$ OAGP},
legend columns=2,
ylabel style={font=\tiny,  yshift=-5mm},
every x tick scale label/.style={at={(rel axis cs:1,0)},anchor=south west,inner sep=1pt},
width=0.22\textwidth,
      height=0.22\textwidth,
ymin=0,
xtick=data,
yticklabel style = {font=\tiny},
xticklabel style = {font=\tiny},
grid=major,
xmajorgrids=false
]
\addplot [fill=gray!10]
	coordinates {
	(200000, 22.018)
	(500000, 29.078)
	(1000000, 39.463)
    (1500000, 60.535)
	(2000000, 89.399)};
\addplot [thick, dash dot]
	coordinates {
	(200000, 25.636)
	(500000, 46.342)
	(1000000, 123.948)
    (1500000, 251.123)
	(2000000, 348.213)};
\addplot [dashed]
	coordinates {
	(200000, 26.761)
	(500000, 29.627)
	(1000000, 46.627)
    (1500000, 64.053)
	(2000000, 98.407)};
\addplot [dotted]
	coordinates {
	(200000, 25.136)
	(500000, 56.342)
	(1000000, 171.948)
    (1500000, 319.131)
	(2000000, 430.213)};
\end{axis}
\end{tikzpicture}
\label{xp:sp-scala}}
\subfloat[]{%
\begin{tikzpicture}
\hspace*{-0.61cm} 
\begin{axis}[
ylabel= Comparisons,
ylabel style={font=\tiny,  yshift=-5mm},
legend style={font=\fontsize{5}{5}\selectfont,
        at={(1.0,1.1)},
        anchor=south east,
        draw=none,
        column sep=1ex,
        },
legend entries={$QueryER$ PPL, $QueryER$ OAGP},
every x tick scale label/.style={at={(rel axis cs:1,0)},anchor=south west,inner sep=1pt},
width=0.22\textwidth,
      height=0.22\textwidth,
yticklabel style = {font=\tiny},
xticklabel style = {font=\tiny},
ymin=0,
xtick=data,
grid=major,
xmajorgrids=false
]
\addplot [fill=gray!10]
	coordinates {
	(200000, 114841)
	(500000, 239799)
	(1000000, 425818)
    (1500000, 609708)
	(2000000, 791348)};
\addplot [dashed]
	coordinates {
	(200000, 42251)
	(500000, 139799)
	(1000000, 344198)
    (1500000, 572566)
	(2000000, 780389)};
\end{axis}
\end{tikzpicture}
\label{xp:sp-scalb}}
\caption{Time and Comparisons for Q9 over \textit{PPL200K-2M} and \textit{OAGP200K-2M}}
\label{xp:sp-scal}
\end{figure}
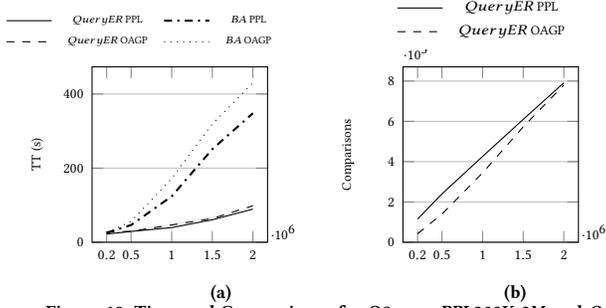

\textbf{Scalability}. Fig.\ref{xp:sp-scala} and Fig.\ref{xp:sp-scalb} evaluate the $TT$ and the executed comparisons of \framework over an increasing dataset size and for a fixed size of $QE_E$. The purpose of this experiment is to demonstrate how \framework scales while increasing only the dataset size. From both metrics, we can safely conclude that \framework scales in a sub-linear fashion while increasing $|E|$. It manages not only to keep the number of comparisons in the same order of magnitude for the whole range of dataset sizes but not even doubling them in cases that the $|E|$ is doubled. The effects of simultaneously increasing both $|E|$ and $|QE_E|$ will be examined in Section \ref{xp:spj}.

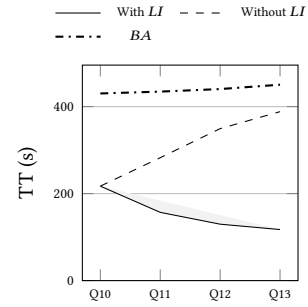
\begin{figure}
\centering
% \subfloat[]{%
\begin{tikzpicture}
\begin{axis}[
ylabel= TT (s),
legend style={font=\fontsize{5}{5}\selectfont,
        at={(1.1,1.05)},
        anchor=south east,
        draw=none,
        column sep=1ex,
        },
legend entries={With $LI$, Without $LI$, $BA$},
legend columns=2,
ylabel style={font=\small,  yshift=-5mm},
every x tick scale label/.style={at={(rel axis cs:1,0)},anchor=south west,inner sep=1pt},
width=0.25\textwidth,
      height=0.25\textwidth,
ymin=0,
xtick=data,
yticklabel style = {font=\tiny},
xticklabel style = {font=\tiny},
symbolic x coords={Q10, Q11, Q12, Q13},
grid=major,
xmajorgrids=false
]
\addplot [fill=gray!10]
	coordinates {
	(Q10, 217.153)
	(Q11, 156.886)
	(Q12, 129.716)
    (Q13, 117.386)};
\addplot [dashed]
	coordinates {
	(Q10, 217.153)
	(Q11, 282.533)
	(Q12, 349.396)
    (Q13, 388.429)};
\addplot [thick,dash dot]
	coordinates {
	(Q10, 430.247)
	(Q11, 434.533)
	(Q12, 440.396)
    (Q13, 450.429)};
\end{axis}
\end{tikzpicture}
% }
\caption{Time for consecutive queries with and without utilizing LI on OAGP2M}
\label{xp:sp-links}
\end{figure}

\textbf{Effects of LI}. Fig.\ref{xp:sp-links} evaluates how the performance of \framework is affected by using the $LI$. As is evident, the $TT$ of each approach diverges from the other with every issued query and is in fact diametrical opposite. The $TT$, of the "Without $LI$" approach is increasing in a sub-linear fashion while approaching the $TT$ of the $BA$. On the contrary, by utilizing the $LI$ and progressively cleaning the $E$, the $TT$ of the "With $LI$" approach is decreasing in the same fashion and approaches 0. 

\textbf{Time breakdown}. Table \ref{tbl:time_breakdown} shows the time breakdown that was computed for $DSD$ and $OAP$ using the query with the highest selectivity ($Q5$). The total time consist of performing (i) \textit{Block-Join}, (ii) \textit{Meta-Blocking}, (iii) \textit{Comparison-Execution}, (iv) \textit{Group-Entities} and (v) other operations (e.g. Table-Scan). The results show that especially on big datasets with high selectivity, \textit{Comparison-Execution} dominates the total time as expected, since distance functions (e.g. Jaro-Winkler) are computationally expensive. If we combine this observation with the observation from Fig.\ref{DSD-comps}, Fig.\ref{oap-comps} and Fig.\ref{oagp2m-comps}, we can safely conclude that the dominating factor of the whole approach is the \textit{Comparison-Execution} since the more comparisons we execute the higher the $TT$ is. 

\textbf{Effects of Meta-Blocking Configurations}. Table \ref{tbl:MB} evaluates the effect of different \textit{Meta-Blocking} configurations on \framework. We executed $Q1$ and $Q5$ on $PPL1M$ and $OAGP1M$ for three different configurations which namely are (i) $ALL$ denoting the combination of all methods, (ii) $BP$+$BF$ denoting the combination of \textit{Block Purging} and \textit{Block Filtering} and (iii) $BP$+$EP$ denoting the combination of \textit{Block Purging} and \textit{Edge Pruning}. As it is observed, among all possible combinations and for both queries, $ALL$ outperforms all other combinations in terms of time but it is outperformed in terms of recall. Since the impact of $ALL$ is not so significant on recall but on the other hand is very high in terms of time, in \framework we used the $ALL$ to sacrifice some recall to enhance performance. All other combinations that are not reported in Table \ref{tbl:MB} took over 1 hour to finish thus they were not considered.
% \vspace{-20pt}
\begin{table}[h]
\centering
\footnotesize
\begin{tabular}{||l|l|l|l|l|l||}
\hline
\textbf{Query}&
\textbf{Method}  &
\textbf{Time (s)} & \textbf{PC} \\ 
\hline
Q1 & ALL & 65.1472 / 70.1352 & 0.918 / 0.832 \\ 
Q1 & BP + BF & 429.207 / 457.32 & 0.996 / 0.987\\
Q1 & BP + EP & $>$ 30 MIN & N/A \\
Q5 & ALL & 106.244 / 112.314 & 0.919 / 0.823 \\ 
Q5 & BP + BF &  980.72 / 802.123 &  0.996 / 0.99\\
Q5 & BP + EP & 
$>$ 30 MIN & N/A \\
% people2m & 15\% & ALL & 185.7158 & 0.911 \\
% people2m & 15\% & BP + BF &
% 1532.72 & 0.997 \\
% people2m & 15\% & BP + EP & > 30 MIN & N/A \\
% people2m & 75\% & ALL &  299.9992 & 0.913\\
% people2m & 75\% & BP + BF &
% > 30 MIN & N/A \\
% people2m & 75\% & BP + EP & > 30 MIN & N/A\\
\hline
\end{tabular}
\caption{\textit{M-B} configurations for $Q1$ and $Q5$ on $PPL1M$ / $OAGP1M$}
\label{tbl:MB}
\end{table}\newline
\vspace{-10pt}

\begin{figure*}[h!]
\centering
\tiny
\subfloat[BA vs NES vs AES TT]{%
\begin{tikzpicture}
\begin{axis}[
ylabel= TT (s),
enlarge x limits=0.5,
width=0.24\textwidth,
       height=0.24\textwidth,
bar width = 4,
ylabel style={font=\tiny,  yshift=-5mm},
xlabel style = {font=\tiny},
yticklabel style = {font=\tiny},
xticklabel style = {font=\tiny},
enlarge x limits=0.7,
legend style={font=\fontsize{4}{5}\selectfont,
                at={(1,1.05)},
                anchor=south east,
                column sep=1ex,
                draw = none
        },
ymin=0,
ybar ,
xtick=data,
legend columns = -1,
symbolic x coords={Q6$_a$, Q7$_a$},
grid=major,
xmajorgrids=false
]

\addplot[fill=gray!10]
	coordinates {
    (Q6$_a$, 421.12)
	(Q7$_a$, 424.341)};
\addplot[fill=gray!40]
	coordinates {
	(Q6$_a$, 180.52)
	(Q7$_a$, 354.664)
	};
\addplot[fill=gray!100]
	coordinates {
    (Q6$_a$, 99.502)
	(Q7$_a$, 292.4188)};

\legend{BA, NES, AES}
\end{axis}
\end{tikzpicture}
\label{nes_vs_aes_time}}
 \hspace{0.5em}
\subfloat[BA vs NES vs AES Exec. Comp.]{%
\begin{tikzpicture}
\begin{axis}[
ylabel= Comparisons,
ylabel style={font=\tiny,  yshift=-7mm},
xlabel style = {font=\tiny},
yticklabel style = {font=\tiny},
xticklabel style = {font=\tiny},
enlarge x limits=0.2,
legend style={font=\fontsize{4}{5}\selectfont,
                at={(1,1.05)},
                anchor=south east,
                column sep=1ex,
                draw = none
        },
enlarge x limits=0.7,
width=0.24\textwidth,
       height=0.24\textwidth,
bar width = 4,
ymin=0,
ybar,
legend columns = 2,
xtick=data,
symbolic x coords={Q6$_a$, Q7$_a$},
grid=major,
xmajorgrids=false
]

\addplot [fill=gray!10]
	coordinates {
	(Q6$_a$, 9409555)
	(Q7$_a$, 9409555)};
\addplot [fill=gray!40]
	coordinates {
	(Q6$_a$, 2274648)
	(Q7$_a$, 7659634)};
\addplot [fill=gray!100]
	coordinates {
	(Q6$_a$, 415540)
	(Q7$_a$, 5707059)};

\legend{BA, NES, AES}
\end{axis}
\end{tikzpicture}
\label{nes_vs_aes_comps}
}
 \hspace{0.5em}
\subfloat[BA vs NES vs AES TT]{%
\begin{tikzpicture}
\begin{axis}[
ylabel= TT (s),
enlarge x limits=0.5,
width=0.24\textwidth,
       height=0.24\textwidth,
bar width = 4,
ylabel style={font=\tiny,  yshift=-5mm},
xlabel style = {font=\tiny},
yticklabel style = {font=\tiny},
xticklabel style = {font=\tiny},
enlarge x limits=0.7,
legend style={font=\fontsize{4}{5}\selectfont,
                at={(1,1.05)},
                anchor=south east,
                column sep=1ex,
                draw = none
        },
ymin=0,
ybar ,
xtick=data,
legend columns = -1,
symbolic x coords={Q6$_b$, Q7$_b$},
grid=major,
xmajorgrids=false
]
\addplot[fill=gray!10]
	coordinates {
	(Q6$_b$, 454.134)
	(Q7$_b$, 457.145)
	};
\addplot[fill=gray!40]
	coordinates {
	(Q6$_b$, 166.313)
	(Q7$_b$, 437.363)
	};
\addplot[fill=gray!100]
	coordinates {
    (Q6$_b$, 62.735)
	(Q7$_b$, 88.476)};
\legend{BA, NES, AES}
\end{axis}
\end{tikzpicture}
\label{nes_vs_aes_time_2}}
 \hspace{0.5em}
\subfloat[BA vs NES vs AES Exec. Comp.]{%
\begin{tikzpicture}
\begin{axis}[
ylabel= Comparisons,
ylabel style={font=\tiny,  yshift=-8mm},
xlabel style = {font=\tiny},
yticklabel style = {font=\tiny},
xticklabel style = {font=\tiny},
enlarge x limits=0.2,
legend style={font=\fontsize{4}{5}\selectfont,
                at={(1,1.05)},
                anchor=south east,
                column sep=1ex,
                draw = none
        },
enlarge x limits=0.7,
width=0.24\textwidth,
       height=0.24\textwidth,
bar width = 4,
ymin=0,
ybar,
legend columns = 2,
xtick=data,
symbolic x coords={Q6$_b$, Q7$_b$},
grid=major,
xmajorgrids=false
]

\addplot [fill=gray!10]
	coordinates {
	(Q6$_b$, 5409555)
	(Q7$_b$, 5409555)};
\addplot [fill=gray!40]
	coordinates {
	(Q6$_b$, 1424506)
	(Q7$_b$, 4718333)};
\addplot [fill=gray!100]
	coordinates {
	(Q6$_b$, 26155)
	(Q7$_b$, 74733)
	};

\legend{BA, NES, AES}
\end{axis}
\end{tikzpicture}
\label{nes_vs_aes_comps_2}
}

\caption{BA vs NES vs AES on TT and E. Comp.}
\label{nes_vs_aes}
\end{figure*}
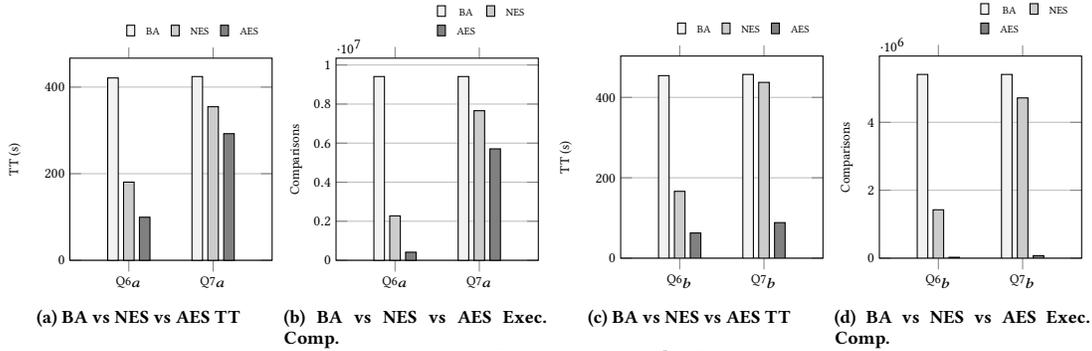

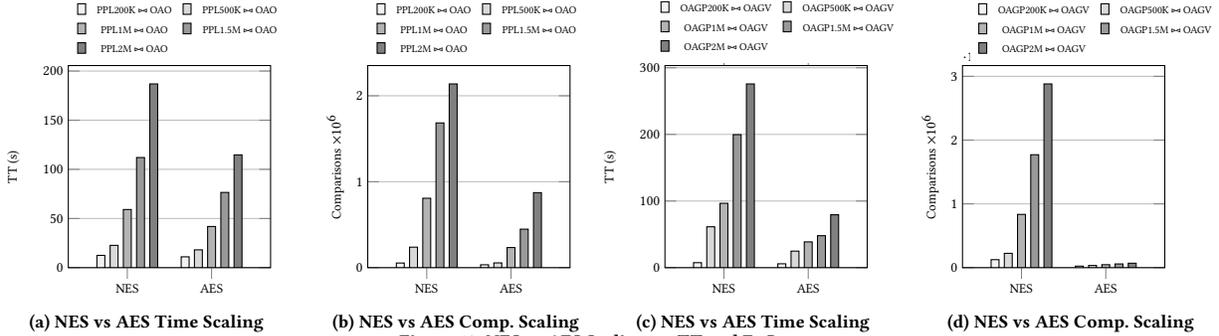
\begin{figure*}[h!]
\centering
\tiny

 \hspace{0.5em}
\subfloat[NES vs AES Time Scaling]{
\begin{tikzpicture}
\begin{axis}[
ylabel= TT (s),
enlarge x limits=0.2,
width=0.24\textwidth,
       height=0.24\textwidth,
bar width = 3,
ylabel style={font=\tiny,  yshift=-5mm},
xlabel style = {font=\tiny},
yticklabel style = {font=\tiny},
xticklabel style = {font=\tiny},
enlarge x limits=0.7,
legend style={font=\fontsize{4}{3}\selectfont,
                at={(1.1,1.0)},
                anchor=south east,
                column sep=1ex,
                draw = none
        },
ymin=0,
ybar,
xtick=data,
symbolic x coords={NES, AES},
grid=major,
legend columns = 2,
xmajorgrids=false
]

\addplot[fill=gray!10]
	coordinates {
	(NES, 12.4864)
	(AES, 11.021)};
\addplot [fill=gray!30]	
	coordinates {
	(NES, 22.684)
	(AES, 18.122)};
\addplot [fill=gray!60]
	coordinates {
	(NES, 59.1312)
	(AES, 41.8686)};
\addplot[fill=gray!80]
    coordinates {
    (NES, 112.0206)
    (AES, 76.5036)
    };
\addplot[fill=gray!100]
	coordinates {
	(NES, 186.8908)
	(AES, 114.726)};

\legend{PPL200K $\bowtie$ OAO, PPL500K $\bowtie$ OAO, PPL1M $\bowtie$ OAO, PPL1.5M $\bowtie$ OAO, PPL2M $\bowtie$ OAO}
\end{axis}
\end{tikzpicture}
\label{nes_vs_aes_time_scale}
}
\hspace{2.5mm}
\subfloat[NES vs AES Comp. Scaling]{
\begin{tikzpicture}
\begin{axis}[
ylabel= Comparisons $\times 10^6$,
enlarge x limits=1,
width=0.24\textwidth,
       height=0.24\textwidth,
bar width = 3,
ylabel style={font=\tiny,  yshift=-8mm},
xlabel style = {font=\tiny},
yticklabel style = {font=\tiny},
xticklabel style = {font=\tiny},
enlarge x limits=0.7,
legend style={font=\fontsize{4}{3}\selectfont,
                at={(1.1,1)},
                anchor=south east,
                column sep=1ex,
                draw = none
        },
ymin=0,
ybar,
xtick=data,
symbolic x coords={NES, AES},
grid=major,
legend columns = 2,
xmajorgrids=false
]

\addplot[fill=gray!10]
	coordinates {
	(NES, 53804)
	(AES, 33452)};
\addplot[fill=gray!30]
	coordinates {
	(NES, 238539)
	(AES, 55489)};
\addplot [fill=gray!60]
	coordinates {
	(NES, 807713)
	(AES, 234187)};
\addplot [fill=gray!80]
	coordinates {
	(NES, 1684257)
	(AES, 448108)};
\addplot[fill=gray!100]
	coordinates {
	(NES, 2138109)
	(AES, 872817)};

\legend{PPL200K $\bowtie$ OAO, PPL500K $\bowtie$ OAO, PPL1M $\bowtie$ OAO, PPL1.5M $\bowtie$ OAO, PPL2M $\bowtie$ OAO}
\end{axis}
\end{tikzpicture}
\label{nes_vs_aes_comps_scale}
}
\subfloat[NES vs AES Time Scaling]{
\begin{tikzpicture}
\begin{axis}[
ylabel= TT (s),
enlarge x limits=0.2,
width=0.24\textwidth,
       height=0.24\textwidth,
bar width = 3,
ylabel style={font=\tiny,  yshift=-5mm},
xlabel style = {font=\tiny,  xshift=-5mm},
yticklabel style = {font=\tiny},
xticklabel style = {font=\tiny},
enlarge x limits=0.7,
legend style={font=\fontsize{4}{3}\selectfont,
                at={(1.2,1.01)},
                anchor=south east,
                column sep=1ex,
                draw = none
        },
ymin=0,
ybar,
xtick=data,
symbolic x coords={NES, AES},
grid=major,
legend columns = 2,
xmajorgrids=false
]
\addplot[fill=gray!10]
	coordinates {
	(NES, 7.4864)
	(AES, 5.808)};
\addplot [fill=gray!30]	
	coordinates {
	(NES, 61.37)
	(AES,  24.797)};
\addplot [fill=gray!60]
	coordinates {
	(NES, 96.594)
	(AES, 38.672)};
\addplot[fill=gray!80]
    coordinates {
    (NES, 199.652)
    (AES, 47.961)
    };
\addplot[fill=gray!100]
	coordinates {
	(NES, 275.768)
	(AES, 79.394)};

\legend{OAGP200K $\bowtie$ OAGV, OAGP500K $\bowtie$ OAGV, OAGP1M $\bowtie$ OAGV, OAGP1.5M $\bowtie$ OAGV, OAGP2M $\bowtie$ OAGV}
\end{axis}
\end{tikzpicture}
\label{nes_vs_aes_time_scale_2}
}
\subfloat[NES vs AES Comp. Scaling]{
\begin{tikzpicture}
\begin{axis}[
ylabel= Comparisons $\times 10^6$,
enlarge x limits=1,
width=0.24\textwidth,
       height=0.24\textwidth,
bar width = 3,
ylabel style={font=\tiny,  yshift=-8mm},
xlabel style = {font=\tiny},
yticklabel style = {font=\tiny},
xticklabel style = {font=\tiny},
enlarge x limits=0.7,
legend style={font=\fontsize{4}{3}\selectfont,
                at={(1.3,1)},
                anchor=south east,
                column sep=1ex,
                draw = none
        },
ymin=0,
ybar,
xtick=data,
symbolic x coords={NES, AES},
grid=major,
legend columns = 2,
xmajorgrids=false
]

\addplot[fill=gray!10]
	coordinates {
	(NES, 125234)
	(AES, 23985)};
\addplot[fill=gray!30]
	coordinates {
	(NES, 224345)
	(AES, 34291)};
\addplot [fill=gray!60]
	coordinates {
	(NES, 835421)
	(AES, 46223)};
\addplot [fill=gray!80]
	coordinates {
	(NES, 1770557)
	(AES, 58198)};
\addplot[fill=gray!100]
	coordinates {
	(NES, 2880725)
	(AES, 69594)};

\legend{OAGP200K $\bowtie$ OAGV, OAGP500K $\bowtie$ OAGV, OAGP1M $\bowtie$ OAGV, OAGP1.5M $\bowtie$ OAGV, OAGP2M $\bowtie$ OAGV}
\end{axis}
\end{tikzpicture}
\label{nes_vs_aes_comps_scale_2}
}
\caption{NES vs AES Scaling on TT and E. Comp.}
\label{nes_vs_aes}
\end{figure*}
\vspace{-2em}
\subsection{\textit{Na\"{i}ve ER Solution} vs \textit{Advanced ER Solution}}\label{xp:spj}
This section demonstrates how the \textit{Naive ER Solution} \textbf{NES}, \textit{Advanced ER Solution} \textbf{AES} and the \textit{Batch Approach} \textbf{BA} perform on SPJ queries and the comparison between them in terms of $TT$ and executed comparisons. Since the $BA$ is not directly applicable in the SPJ class of queries, in all cases, both tables were deduplicated prior to the $Join$ operation and the accumulation of the individual metrics is reported.
Fig.\ref{nes_vs_aes_time},\subref{nes_vs_aes_time_2} and Fig.\ref{nes_vs_aes_comps},\subref{nes_vs_aes_comps_2} show the $TT$ and executed comparisons for $AES$, $NES$ and $BA$ for $Q6_{a,b}$ and $Q7_{a,b}$ respectively. As expected $AES$ outperforms both $NES$ and $BA$ in terms of $TT$ and executed comparisons. This is due to its awareness of the query's statistics which enables it to perform the optimal join ordering but most importantly to decide the best ER operators placement as explained in Section \ref{Query Planning}, resulting in savings in the executed comparisons. \silence{Once more, we can observe that the executed comparisons are strongly correlated and are analogous to the time.} The difference that $NES$ and $BA$ exhibit in $Q7_a$ and $Q7_b$, where the selectivity is ~$75\%$ and $100\%$ respectively, decreases. This is because the parts of data that influence the query’s answer tend to be equal to the size of the whole dataset, just like in section \ref{SP-EXPS}. On the contrary, the fact that, $AES$ is heavily based on cleaning the table that yields the highest number of comparisons first, is clearly proven by its difference from $NES$ and $BA$ in the aforementioned queries. While table sizes tend to be equal, the performance of $AES$ is only affected by the the join-percentage.

An interesting observation in Fig.\ref{nes_vs_aes_time_2},\subref{nes_vs_aes_comps_2}, is that the $TT$ of $AES$ unlike the one of $NES$ is not so strongly correlated analogously to the executed comparisons. This is because, the join-percentage of these two tables is small (5\%) and thus the $QE_E$ of $OAGP2M$ that is formed from the entities that join with the $\mathcal{DR}_E$ of $OAGV$ is small too. This is important because, as stated before, the number of executed comparisons is analogous to the size of the $QE_E$ and on this instance the $TT$ is dominated by the blocking/meta-blocking operations. More specifically, the time breakdown for the different deduplicate operations is as follows: Blocking 10\%, Block-join 3\%, Block Purging 0.5\%, Block Filtering 0.5, Edge Pruning 82\%, Comparison Execution 4 \%. The fact that these findings are contradictory with the results of table \ref{tbl:time_breakdown}, indicates that deduplicating the table that yields the least comparisons first is even more critical for the performance in such cases. 

Fig.\ref{nes_vs_aes_time_scale},\subref{nes_vs_aes_comps_scale} evaluate the $TT$ and the executed comparisons for $AES$ and $NES$ for $Q8_a$ which performs joins between $PPL200K-2M$ and $OAO$ for fixed selectivity $15\%$ and $100\%$ respectively, thus examining the scalability of both approaches over an increasing $|E|$ with an also increasing $|QE_E|$. Again, $AES$ outperforms $NES$ but  both approaches scale in a sub-linear fashion over the increasing dataset size. This fact can be more easily observed in Fig.\ref{nes_vs_aes_comps_scale} where we can see that even if doubling the dataset size, the order of magnitude for the comparisons remains the same. A more interesting observation, is that in this case not only the dataset size is increasing but also the original number of the entities ($|QE_E|$) for each dataset and since the number of the comparisons is depending heavily on $|QE_E|$ this could have severe consequences in the performance. Yet, the scaling remains sub-linear.

Fig.\ref{nes_vs_aes_time_scale_2},\subref{nes_vs_aes_comps_scale_2} evaluate the $TT$ and the executed comparisons for $AES$ and $NES$ for $Q8_b$ which performs joins between $OAGP200K-2M$ and $OAGV$. The results of these experiments are comparable to those of Fig.\ref{nes_vs_aes_time_scale},\subref{nes_vs_aes_comps_scale}. Here again, like on Fig.\ref{nes_vs_aes_time_2},\subref{nes_vs_aes_comps_2} the Executed Comparisons of AES are not analogous to the $TT$ and this again has to do with the low join-percentage of the two tables.
\vspace{-2pt}
\subsection{Summary} \label{summary}
Our thorough experimental study leads to several observations about the performance of our approach.\par
First, the lower the selectivity ($|QE_E|$) of a given query, the more the \framework outperforms $BA$. In the $SPJ$ class, this translates to only deduplicating the entities that join between the two tables, thus, making \framework a scalable solution especially when the join-percentage is low and the sizes of the tables present great inequality. For instance, imagine if an analyst deduplicated a huge table first only to find out that a very small percentage of it joins with the other table that is significantly smaller. A \textit{Batch Approach} would be unable to handle such case efficiently. These facts further support our motivation and prove that \framework is well suited for data exploration and analysis workflows.\par
Another interesting observation is that in cases of very small $|QE_E|$ the comparison-execution does not dominate the total execution time $TT$. In such cases, $TT$ is dominated by the Meta-Blocking, especially by $EP$. Initially, that finding seems to be contradictory with the results of Table \ref{tbl:time_breakdown} but with a closer look at Table \ref{tbl:MB} it can be observed that the $ALL$ is still the most efficient configuration.\par
On the whole, \framework has been evaluated in several real-world and synthetic datasets. It scales much better than $BA$ with respect to $TT$ and executed comparisons for a given query. Across all our experiments, the levels of $PC$ never dropped below $82\%$ with a mean of $91\%$. Finally, the addition of the $LI$ improves the overall performance drastically by enabling the deduplication in a progressive manner. This approach can be very useful also in data exploration and analysis scenarios, where the analyst explores the datasets with consecutive queries that are often overlapping.
\section{Conclusions and Future Work} \label{Conclusions}

 In this work, we studied the problem of integrating Entity Resolution into the traditional query processing in the context of SPJ SQL queries. We developed \framework, a novel framework that enables the performance of analysis-aware entity resolution over dirty data with duplicate entries with minimum pre-processing time and no preparation overhead. \silence{overhead}We evaluated \framework over both real and synthetic datasets and showed that it scales in a sub-linear fashion while achieving high levels of recall. This work revealed several new directions for future research. The extension of \framework to cope with other classes of queries (e.g. aggregation  and analytical queries) is an interesting direction for future work. Another direction is the integration of different blocking and entity matching methods in our analysis framework and their comparative evaluation w.r.t. efficiency and effectiveness. Finally, we plan to scale out our implementation on a distributed environment.
 
 \textbf{Acknowledgements.} The \framework project (1614) has been funded by the Hellenic Foundation for Research and Innovation (ELIDEK) and by the General Secretariat for Research and Technology (GSRT).
\bibliographystyle{ACM-Reference-Format}
\bibliography{output}
\end{document}